\documentclass[%
reprint, 
superscriptaddress,
showpacs,preprintnumbers,
nofootinbib,
 amsmath,amssymb,
 aps,
pre,
]{revtex4-1}

\usepackage[utf8]{inputenc}
\usepackage[colorlinks=true,urlcolor=blue,linkcolor=blue,citecolor=blue]{hyperref}
\usepackage[T1]{fontenc}
\usepackage{lmodern}

\usepackage{graphicx}
\usepackage{dcolumn}
\usepackage{bm}
\usepackage[usenames, dvipsnames]{color}
\usepackage{siunitx}
\begin{document}

\preprint{APS/123-QED}

\title{Multiscale modeling of polycrystalline graphene: A comparison of structure and defect energies of realistic samples from phase field crystal models}

\author{Petri Hirvonen}
\email[email: ]{petri.hirvonen@aalto.fi}
\affiliation{COMP Centre of Excellence, Department of Applied Physics, Aalto University School of Science, P.O. Box 11000, FIN-00076 Aalto, Espoo, Finland}

\author{Mikko M. Ervasti}
\affiliation{COMP Centre of Excellence, Department of Applied Physics, Aalto University School of Science, P.O. Box 11000, FIN-00076 Aalto, Espoo, Finland}

\author{Zheyong Fan}
\affiliation{COMP Centre of Excellence, Department of Applied Physics, Aalto University School of Science, P.O. Box 11000, FIN-00076 Aalto, Espoo, Finland}

\author{Morteza Jalalvand}
\affiliation{Department of Physics, University of Tehran, Tehran 14395-547, Iran}
\affiliation{Department of Physics, Institute for Advanced Studies in Basic Sciences (IASBS), Zanjan 45137-66731, Iran}

\author{Matthew Seymour}
\affiliation{Department of Physics, McGill University, 3600 University Street, Montreal Quebec, H3A2T8}

\author{S. Mehdi Vaez Allaei}
\affiliation{Department of Physics, University of Tehran, Tehran 14395-547, Iran}

\author{Nikolas Provatas}
\affiliation{Department of Physics, McGill University, 3600 University Street, Montreal Quebec, H3A2T8}

\author{Ari Harju}
\affiliation{COMP Centre of Excellence, Department of Applied Physics, Aalto University School of Science, P.O. Box 11000, FIN-00076 Aalto, Espoo, Finland}

\author{Ken R. Elder}
\affiliation{Department of Physics, Oakland University, Rochester, Michigan 48309, USA}

\author{Tapio Ala-Nissila}
\affiliation{COMP Centre of Excellence, Department of Applied Physics, Aalto University School of Science, P.O. Box 11000, FIN-00076 Aalto, Espoo, Finland}
\affiliation{Department of Physics, Brown University,
P.O. Box 1843, Providence, Rhode Island 02912, USA}


\date{\today}

\begin{abstract}
We extend the phase field crystal (PFC) framework to quantitative modeling of polycrystalline graphene.
PFC modeling is a powerful multiscale method for 
finding the ground state configurations of large realistic samples that can be further 
used to study their mechanical, thermal or electronic properties.
By fitting to quantum-mechanical density functional theory (DFT) calculations, 
we show that the PFC approach is able to predict realistic formation energies and defect
structures of grain boundaries.
We provide an in-depth comparison of the formation energies between PFC, DFT and molecular dynamics (MD) calculations.
The DFT and MD calculations are initialized using 
atomic configurations extracted from PFC ground states.
Finally, we use the PFC approach 
to explicitly construct large realistic polycrystalline samples and 
characterize their properties using MD relaxation to demonstrate their quality.

\end{abstract}

\maketitle


\section{Introduction}

Graphene is an intensely studied material due to its remarkable mechanical strength, and extraordinary thermal and electrical conductivities \cite{Novoselov_2004, Lee_2008, Balandin_2008, Castro_Neto_2009}.
Graphene-based devices and interconnects often require high quality samples, whereas large graphene patches typically grown by the industry-standard chemical vapor deposition (CVD) result in polycrystalline structures \cite{ref-quilts, ref-kim}.
Polycrystalline graphene is a quilt consisting of pristine graphene domains in various orientations that are separated by grain boundary defects comprised of dislocations that accommodate the lattice mismatch between neighboring grains.
It is the grain boundaries that largely determine the properties of the material, such as out-of-plane relaxation, weakening its mechanical strength at low-angle tilt boundaries, and altering the electronic structure and transport properties \cite{ref-pringles, Grantab_2010, ref-yazyev-louie, Yazyev_2010, ref-valley, ref-QHE1, ref-QHE2, ref-QHE3, ref-graphene-defects}.
Many of the features of the grain boundaries stem from the early stages of formation, where the graphene grains nucleate in certain orientations that are partly determined by the substrate. 
This results in a rich variety of possible grain tilt angles and grain boundary topologies, each having their own characteristic properties. These defected structures have been analyzed in several recent works both experimentally and theoretically \cite{ref-graphene-review-1, ref-graphene-review-2}.


Modeling realistic systems of polycrystalline graphene has remained a challenge due to the multiple length and time scales involved.
Of the conventional methods, quantum mechanical density functional theory (DFT) is limited to small sample sizes with a few thousand atoms at best, whereas the time scales of tracing atomic vibrations in molecular dynamics (MD) simulations are too short to capture dislocation dynamics.
Constructing model systems with grains and grain boundaries has, therefore, typically been approached as a multi-step process, using for instance cut and paste, iterative grain growth, thermalization and cooling of the grain boundaries by applying local relaxation, and probing stability by adding additional atoms \cite{ref-liu, ref-van-tuan}. For constructing symmetric grains, the coincidence site lattice (CSL) theory can be applied \cite{ref-yazyev-louie, ref-carlsson}.
In the general case, however, there still are obvious problems in the construction of realistic samples such as determining how many carbon atoms are needed at the grain boundary, and whether the low-stress ground state configuration has been reached. Furthermore, one commonly restricts to the 5|7 dislocation defects with adjacent pentagon-heptagon pairs in the graphene backbone that have been seen in experiments using transmission electron microscopy techniques \cite{ref-quilts, ref-kim}. However, in some tilt angles and conditions there could be other interesting defect types present, such as 5|8|7 defects that have been shown to have finite spin moments \cite{ref-dangling}.
Other polygons and more complex chains are possible in principle as well, but the number of structural permutations corresponding to reasonable grain boundaries is too large to be sampled by conventional microscopic computational methods.

Our solution to multiscale modeling of polycrystalline graphene is to apply phase field crystal (PFC) models \cite{ref-pfc-debut}. They are ideally suited to deal with large system sizes required by the polycrystalline nature of graphene. Namely, PFC is a continuum approach to microstructure evolution and elastoplasticity in crystalline materials. It models a time-averaged atomic number density field over long, diffusive time scales, while retaining atomic-level spatial resolution.
Due to the relative simplicity of PFC models, and the numerically convenient smoothness of the density fields, mesoscopic length scales are easily attained. Therefore, PFC models can be used to construct even large realistic systems without a priori knowledge of the atomic positions. The multiscale characteristics of the PFC framework allow access to new modeling regimes that fall beyond the reach of conventional techniques \cite{ref-pfc-debut, ref-pfc-2004}.

In this work, we carry out a thorough evaluation of four different two-dimensional PFC models by studying graphene grain boundary structures and energies at varying tilt angles and with different defect types. Such structure and energetics calculations have been performed previously with
MD \cite{ref-carlsson, ref-pringles, ref-liu}, DFT \cite{ref-yazyev-louie, ref-zhang}, density-functional tight-binding \cite{ref-malola}, and with a combination of several methods \cite{ref-dangling, ref-graphene-failure}. While PFC approaches have already been used to study certain topological features of graphene \cite{ref-ruga, ref-matt}, our focus is to find a PFC model that is suited for quantitative modeling and evolution of large multi-grain graphene systems. 
In order to determine the absolute energy scale, the PFC models are fitted to 
DFT by matching the grain boundary formation energies at small tilt angles.
Then by using the same PFC constructed initial atomic geometries, we present 
an extensive comparison of the formation energies of various grain boundaries relaxed and evaluated using PFC, 
and DFT and MD in two and three dimensions.

To further validate the use of PFC models to study polycrystalline graphene, we examine in detail ground state configurations of grain boundaries and the distributions of different defect or dislocation types produced by the PFC models. While 5|7 grain boundaries are the most prominent, one of the PFC models produces a rich variety of alternate dislocation types in certain tilt angle regimes. Furthermore, we explicitly show that the PFC models can be used to construct large and low-stress polycrystalline graphene systems up to
hundreds of nanometers in linear size for further mechanical, thermal or electronic transport calculations.
Here, we characterize the formation energies of such realistic systems of varying size.

This paper is organized as follows.
Section II introduces the DFT, MD and PFC models.
In Section III, the PFC models are fitted to DFT, and are used to study both the formation energy and topology of grain boundaries.
In Section IV, we construct large polycrystalline samples and demonstrate their quality by characterization of their properties.
Section V presents our summary and conclusions.

\section{Methods}

\subsection{Quantum mechanical density functional theory calculations}
\label{sec-DFT}

The most fundamental practical method for calculating materials properties is based on solving the Schrodinger equation for
the system under study. In the Born-Oppenheimer approximation, the electronic structure and the nuclear configuration are solved separately. Density functional theory solves the quantum-mechanical electronic structure of a material, after which the atomic geometry can be relaxed using the forces evaluated from the DFT total energy gradients. While this constitutes a highly accurate quantum
mechanical description, the system sizes are severly limited by the computational cost.

The DFT calculations here were performed by initializing the systems in 2D by PFC. To guarantee quantitative accuracy
of the DFT calculations for graphene, we used the all-electron FHI-aims package \cite{Blum_2009}. It uses numerical atom-centered basis functions for each atom type. The default \textit{light} basis sets were employed together with the GGA-PBE functional \cite{Perdew_1996}. During the course of the calculation, the self-consistent cycle was considered converged if, among other things, the total energy had converged up to $10^{-6}$ eV between consecutive iterations. The atom geometries were relaxed in each case until the forces acting on the atoms were smaller than $10^{-2}$ eV/\AA.

\subsection{Molecular dynamics calculations}
\label{sec-MD}

Molecular dynamics (MD) methods comprise further coarse-graining as compared to DFT by replacing the electronic structure with effective interatomic potentials. As a consequence, computational complexity is reduced and very large systems with millions of atoms can be currently handled. However, tracing atomic vibrations at femtosecond time-scales becomes the stumbling block. That is, processes such as microstructure formation and evolution that occur over long, diffusive time scales cannot usually be addressed. Constructing large polycrystalline samples with low stress also becomes a difficult task, since the relaxed structure can be nontrivial and hence not known a priori.

In this work, MD was used to calculate formation energies of grain boundaries and to characterize the properties of large polycrystalline samples constructed using PFC. The grain boundary formation energies were evaluated by MD calculations using the 
Large-scale Atomic/Molecular Massively Parallel Simulator (LAMMPS) software \cite{lammps-plimpton}. Two potentials were used to define the interactions of carbon atoms: the adaptive intermolecular reactive empirical bond order (AIREBO) potential \cite{airebo-stuart} and the Tersoff potential \cite{tersoff}. We employed the parameters provided by S. J. Stuart {\it et al.} 
\cite{airebo-stuart} for the AIREBO potential and the parameters provided by J. Tersoff \cite{tersoff-param} for the Tersoff potential. The Polak-Ribiere version of the conjugate gradient algorithm \cite{polak-ribiere-cg} was used in all of the minimizations. All minimizations were carried out until one of these criteria was met: The energy change between two successive iterations is less than $10^{-6}$ times its magnitude, or length of the global force vector is less than $10^{-6}$ eV/\AA.

The formation energies of grain boundaries were also evaluated using a MD code implemented fully on graphics process units (GPUs) \cite{fan2013,fan2015}, which can be two orders of magnitude faster than a serial code for large systems. This code uses the Tersoff potential \cite{tersoff}, but with optimized parameters provided by L. Lindsay and D. A. Broido \cite{lindsay2010}, which are better suited for modeling graphene. In the calculation of the grain boundary formation energies, we performed MD simulations at a low temperature of 1 K with a total simulation time of 100 ps, where the systems become fully relaxed.

The relative efficiency of the Tersoff potential as compared to the AIREBO potential and its acceleration by GPUs allowed us to simulate large-scale polycrystalline graphene samples with long simulation times. Here, a room temperature of 300 K was chosen and the in-plane stress was required to be around zero.
All simulations of polycrystalline graphene samples were performed up to 1000 ps to ensure full convergence of out-of-plane deformations. In all the MD simulations with the GPU code, we adopted the Verlet-velocity integration method with a time step of 1 fs.

\subsection{Phase field crystal models}
\label{sec-pfc-calculations}

PFC is a continuum approach that models crystalline matter via a classical density field, $\psi = \psi\left(\boldsymbol{r}\right)$ \cite{ref-pfc-debut}. The density field is governed by a free energy functional, $F = F\left[\psi\right]$, that is chosen to be minimized by a periodic solution to $\psi$. The relaxed configuration corresponding to a particular initial state can be solved via energy minimization. Details of the functional determine the symmetries in the ground state that can be matched with the desired crystal structure.

The standard relaxational dynamics for PFC capture dynamics on diffusive time scales only. Thereby the atomic vibrations captured by MD are effectively coarse-grained into time-averaged smooth peaks in $\psi$ describing the lattice. The main focus of this work is finding the lowest-energy states that contain grain boundaries, not on the dynamics of the formation of such states. In this instance it is not necessary to use the traditional conserved relaxational dynamics. For this reason a variety of more advanced approaches were used to obtain the lowest energy structures as discussed in detail in Appendices \ref{sec-PFC-further} and \ref{sec-ini-rela}.

In the PFC approach, atomic resolution is retained while the smoothness of $\psi$ facilitates numerical modeling of systems with millions of atoms. The PFC description of matter neglects some microscopic details, such as atomic vibrations and vacancies, but it resolves the length and time scale limitations of DFT and MD, respectively \cite{ref-pfc-debut, ref-pfc-2004}.

We investigate here the suitability of four PFC variants for modeling of polycrystalline graphene samples: the one-mode model (PFC1), the amplitude model (APFC), the three-mode model (PFC3) \cite{ref-three-mode} and the structural model (XPFC) \cite{ref-matt}. For the convenience of the reader, more comprehensive details of these models, such as model parameter choices, methods of relaxation, etc., are provided in Appendices \ref{sec-PFC-further} and \ref{sec-ini-rela}.

PFC1 is the standard PFC model \cite{ref-pfc-debut}, but instead of the close-packed triangular lattice formed by its density field maxima, we relate the hexagonal arrangement of density field minima to the atomic positions. In practice, we choose model parameters to invert the density field to yield a hexagonal (triangular) set of maxima (minima). The PFC1 free energy functional is given by
\begin{equation}
\label{eq-PFC1}
F_1=c_1\int d \boldsymbol{r} \left(\frac{\psi\mathcal{L}_1\psi}{2}+\frac{\tau\psi^3}{3}+\frac{\psi^4}{4}\right),
\end{equation}
where
\begin{equation}
\label{eq-L1}
\mathcal{L}_1=\epsilon+\left(q_0^2+\nabla^2\right)^2,
\end{equation}
is a rotation-invariant Hamiltonian describing non-local contributions. The parameter $\epsilon$ is related to temperature, $q_0$ controls the equilibrium lattice constant and $\tau$ sets the average density. The coefficient $c_1$ allows controlling the energy scale of the model.

APFC is an amplitude expanded reformulation of PFC1 \cite{ref-goldenfeld} where the density field is replaced by three smooth, complex-valued amplitude fields, $\eta_j$, for increased numerical performance. The APFC functional is written
\begin{equation}
\label{eq-APFC}
\begin{split}
F_A=c_A\int d \boldsymbol{r} \left(\vphantom{\left(\prod_{j=0}^2\right)}
\frac{\Delta B}{2} A^2+\frac{3v}{4} A^4\right. \\
\left.-2t\left(\prod_{j=0}^2 \eta_j+c.c.\right)\right. \\
\left.+\sum_{j=0}^2 \left(B^x \left|\mathcal{G}_j \eta_j\right|^2-\frac{3v}{2}\left|\eta_j\right|^4\right)
\vphantom{\left(\prod_{j=0}^2\right)}\right),
\end{split}
\end{equation}
where
\begin{equation}
\label{eq-APFC-dB}
\Delta B=B^l-B^x,
\end{equation}
\begin{equation}
\mathcal{G}_j=\nabla^2+2\imath\boldsymbol{g_j}\cdot\nabla,
\end{equation}
and
\begin{equation}
A^2=2\sum_j\left|\eta_j\right|^2.
\end{equation}
The parameters $B^l$ and $B^x$ are related to the compressibility of the liquid state and the elastic moduli of the crystalline state, respectively, whereas the magnitude of the amplitudes and the liquid-solid miscibility gap depend on the choice of $t$ and $v$ \cite{ref-vili}. These parameters were chosen to conform with those of PFC1, see Appendix \ref{sec-PFC-further}. The complex conjugate is denoted by $c.c.$, the imaginary unit by $\imath$, and the lowest-mode set of reciprocal lattice vectors by $\boldsymbol{g_j}$, where $\boldsymbol{g_0}, \boldsymbol{g_1}, \boldsymbol{g_2} = \left(-\sqrt{3}/2, -1/2\right), \left(0, 1\right), \left(\sqrt{3}/2, -1/2\right)$. The coefficient $c_A$ controls the energy scale of the model. The real-space density field can be reconstructed from the complex amplitudes as
\begin{equation}
\label{eq-reconstruction}
\psi\left(\boldsymbol{r}\right)=\sum_j \eta_je^{\imath\boldsymbol{g_j}\cdot\boldsymbol{r}}+c.c.
\end{equation}
It should be noted that this model is limited to relatively small orientational mismatch between neighboring crystals, see Reference \cite{ref-nik} for details.

PFC3 is a generalization of PFC1 that incorporates not just one, but three controlled length scales, or modes. The free energy reads
\begin{equation}
\label{eq-PFC3}
F_3=c_3\int d \boldsymbol{r} \left(\frac{\psi\mathcal{L}_3\psi}{2}+\frac{\psi^4}{4}+\mu\psi\right),
\end{equation}
with
\begin{equation}
\label{eq-L3}
\begin{split}
\mathcal{L}_3=\epsilon+\lambda\left(b_0+\left(q_0^2+\nabla^2\right)^2\right)\left(b_1+\left(q_1^2+\nabla^2\right)^2\right) \\ \times\left(b_2+\left(q_2^2+\nabla^2\right)^2\right),
\end{split}
\end{equation}
where a chemical potential term replaces the third-order term and assumes its role in fixing the average density to a constant value. The third-order term is often omitted from PFC formulations and this choice is argued further in Reference \cite{ref-extended}. The parameters $\lambda$, $b_0$, $b_1$ and $b_2$ weight the competing modes controlled by $q_0$, $q_1$ and $q_2$. Again, the coefficient $c_3$ controls the energy scale.

The XPFC model has a free energy that can be expressed as a sum of three contributions: ideal free energy, $F_{.}$, two-point interactions, $F_{-}$, and three-point interactions, $F_{\Delta}$,
\begin{equation}
\label{eq-XPFC}
F_X = c_X\left(F_{.}+F_{-}+F_{\Delta}\right),
\end{equation}
where $c_X$ sets the energy scale. The ideal free energy is given by
\begin{equation}
\label{eq-XPFC-id}
F_{.} = \int d\boldsymbol{r} \left(\frac{\psi^2}{2}-\eta\frac{\psi^3}{6}+\chi\frac{\psi^4}{12}+\mu\psi\right),
\end{equation}
where $\eta$ and $\chi$ are phenomenological parameters and $\mu$ is 
again the chemical potential. The two-point term is given by
\begin{equation}
\label{eq-XPFC-2p}
F_- = -\frac{1}{2} \int d\boldsymbol{r} \left(\psi \mathcal{F}^{-1}\left\lbrace\hat{C}_2\hat{\psi}\right\rbrace\right),
\end{equation}
where the carets and $\mathcal{F}^{\left(-1\right)}$ denote (inverse) Fourier transforms, and
\begin{equation}
\label{eq-XPFC-C2}
\hat{C}_2 = -2R\frac{J_1\left(r_0 k\right)}{r_0 k},
\end{equation}
where $R$ and $r_0$ set the magnitude and range of the interaction, respectively, $J_1$ is a 
Bessel function of the first kind and $k = \vert \boldsymbol{k} \vert$ vector. The three-point term reads
\begin{equation} 
\label{eq-XPFC-3p}
F_{\Delta} = -\frac{1}{3} \int d\boldsymbol{r} \left(\psi \sum_{i = 1}^2 \left(\mathcal{F}^{-1}\left\lbrace\hat{C}_s^{\left(i\right)}\hat{\psi}\right\rbrace\right)^2\right),
\end{equation}
where
\begin{equation} 
\label{eq-XPFC-Cs12}
\hat{C}_{s}^{(1)} = X {\imath}^m \cos{(m\theta_k)} J_m (k a_0),
\end{equation}
and
\begin{equation}
\label{eq-XPFC-Cs2}
\hat{C}_s^{(2)} = X {\imath}^m \sin{(m\theta_k)} J_m(k a_0).
\end{equation}
Here, $X$ sets the interaction strength, $\imath$ is the imaginary unit, $m = 3$ indicates the three-fold rotational symmetry that is desired here, $\theta_k$ is the polar coordinate angle in Fourier space, and $a_0$ controls the lattice constant.

We apply these models in two dimensions where a number of different phases can be produced depending on the model in question and the set of model parameters employed. These phase diagrams also indicate the possible coexistences and transitions between neighboring phases. We fixed the parameters of each model---excluding the energy scale coefficients $c_1, c_A, c_3$ and $c_X$---well within the hexagonal phase of each model 
to ensure good stability of hexagonal structures. The parameter values and the 
phase diagrams are given in Appendix \ref{sec-PFC-further} and in References \cite{ref-nik, ref-three-mode, ref-matt}, respectively.

\begin{figure}
\includegraphics[width=0.48\textwidth]{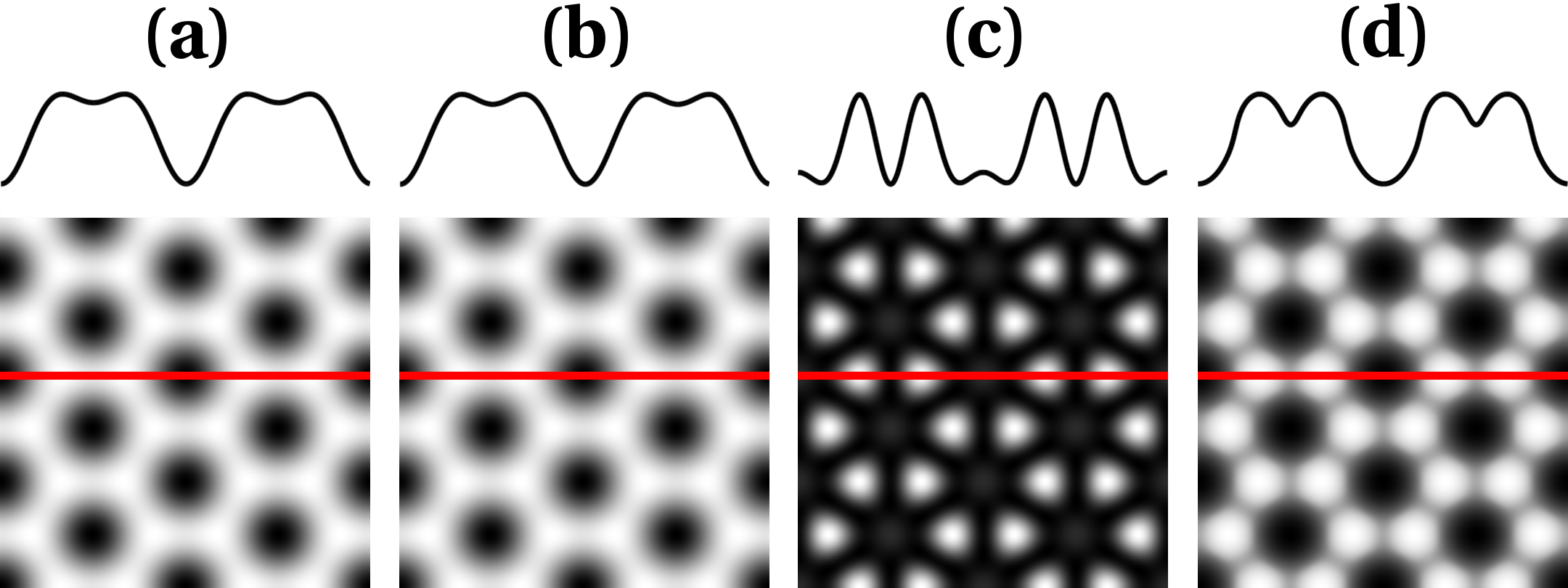}
\caption{(Color online) The appearance of PFC density fields in equilibrium: (a) PFC1, (b) APFC, (c) PFC3 and (d) XPFC. All density fields have been mapped linearly to grayscale values with the maxima (minima) appearing as white (black). Density profiles along the red lines intersecting local maxima and minima are shown on top in arbitrary units.}
\label{fig-fields}
\end{figure}

Figure \ref{fig-fields} showcases the appearance of the density fields of the four models in equilibrium. Density profiles along a straight path coinciding with local maxima and minima are also outlined above the panels. The PFC1 density field, shown in Figure \ref{fig-fields} (a), has a honeycomb mesh-like appearance with weak maxima and pronounced minima. The APFC density field reconstructed from the complex amplitudes appears identical to that of PFC1, see Figure \ref{fig-fields} (b). The respective density profiles also appear identical. Amidst the prominent PFC3 primary maxima, one may notice weak secondary maxima in Figure \ref{fig-fields} (c) that are likely to contribute to the richness of defect structures observed for this model. The corresponding density profile reveals these secondary maxima more clearly. The XPFC density field in Figure \ref{fig-fields} (d) is intermediate between PFC1 and PFC3 with distinct, yet somewhat interconnected maxima.

\section{Grain boundaries}
\label{sec-results}

\subsection{Construction of grain boundaries}

\label{sec-construction}

Extensive calculations of graphene grain boundary topologies and formation energies were performed to benchmark the four PFC models. These results are compared against DFT and MD calculations of identical grain boundaries from both the present and previous works. To simplify the analysis, we considered only symmetrically tilted grain boundaries in systems that were both free-standing and planar. Free-standing systems were treated to facilitate comparison to previous theoretical works and two-dimensionality is a limitation of the PFC models investigated. On the other hand, graphene is typically grown on a substrate \cite{ref-rise-of-graphene, ref-quilts, ref-kim} forcing a planar atomic configuration. Periodic boundary conditions were employed to eliminate edge effects.

\begin{figure}
\includegraphics[width=0.36\textwidth]{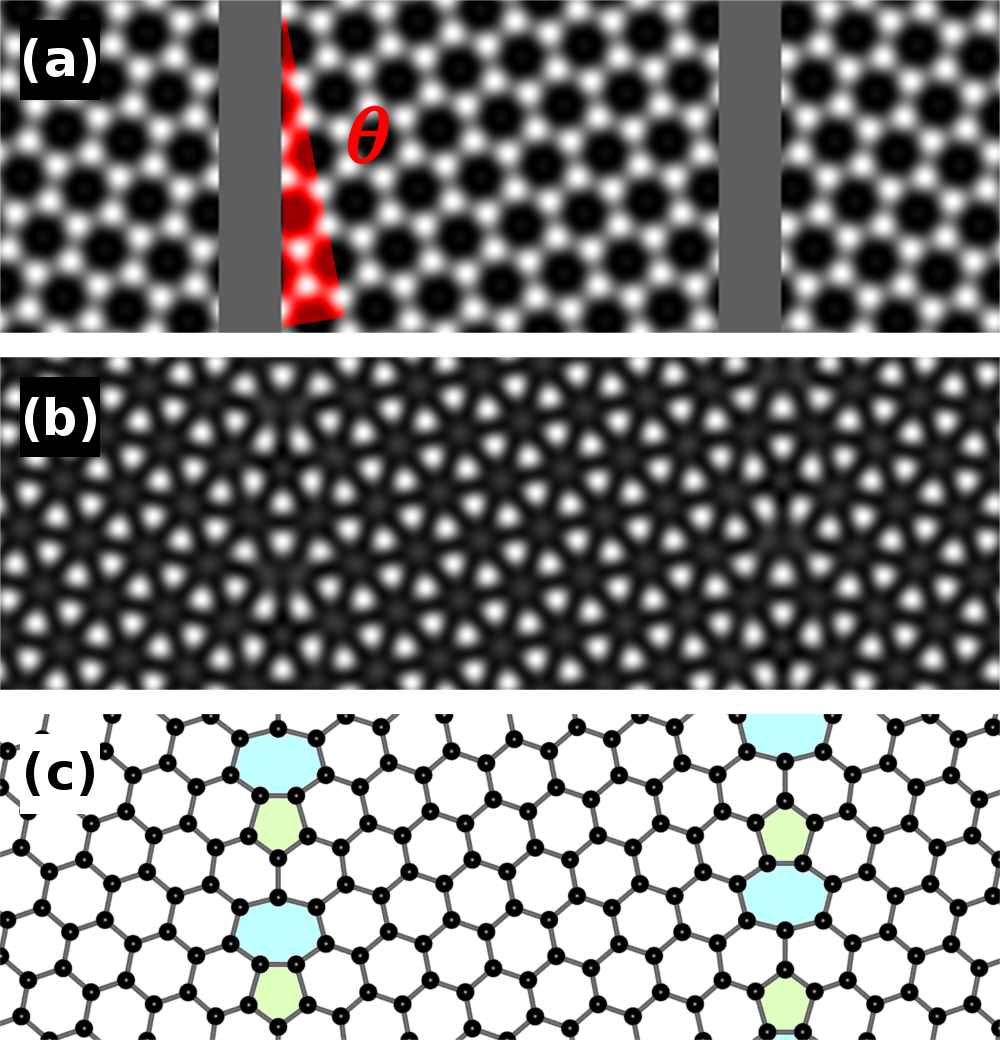}
\caption{(Color online) The layout for grain boundary calculations. (a) The initial state obtained from the one-mode approximation with the grain boundaries in a constant density state (gray vertical bands). Half the tilt angle is highlighted (red wedge). (b) The corresponding relaxed PFC3 configuration and this converted into (c) an atomic configuration. The relaxed grain boundaries are comprised of 5|7 dislocations where the pentagons and heptagons have been colored in green and blue, respectively. Width of the depicted system is $w \approx 4$ nm and its tilt angle $2\theta \approx 21.8^\circ$.}
\label{fig-layout}
\end{figure}

A bicrystalline layout was used for the grain boundary calculations because of periodic boundary conditions. Figure \ref{fig-layout} demonstrates a bicrystal with two grains and two grain boundaries. The tilt angle, $2\theta$, is the difference in crystallographic orientation between the bicrystal halves rotated by $\pm\theta$, see Figure \ref{fig-layout} (a). We take $2\theta\rightarrow 0^\circ$ and $2\theta\rightarrow 60^\circ$ to correspond to armchair and zigzag grain boundaries, respectively, and refer collectively to both limits as small tilt angles.

The symmetrically tilted, hexagonal crystals were constructed into a rectangular, two-dimensional computational unit cell. The initial hexagonal state shown in Figure \ref{fig-layout} (a) was obtained using the one-mode approximation \cite{ref-pfc-2004}
\begin{equation}
\begin{split}
\psi(x,y) = \cos{\left(qx\right)}\cos{\left(qy/\sqrt{3}\right)}
\\
-\cos{\left(2qy/\sqrt{3}\right)}/2,
\end{split}
\end{equation}
where $2\pi/q$ is the lattice constant. While rotating $\psi$ is trivial, the rotated equilibrium state in APFC is given by
\begin{equation}
\eta_j\left(\boldsymbol{r}, \theta\right)=\phi_{eq}\exp{\left(\imath\left(\boldsymbol{g_j}\left(\theta\right)-\boldsymbol{g_j}\left(0\right)\right)\cdot\boldsymbol{r}\right)},
\end{equation}
where
\begin{equation}
\phi_{eq}=\frac{t-\sqrt{t^2-15v\left(B^l-B^x\right)}}{15v},
\end{equation}
and
\begin{equation}
\boldsymbol{g_j}\left(\theta\right)=
\left(\begin{matrix}
 \cos{\left(\theta\right)} & -\sin{\left(\theta\right)} \\
 \sin{\left(\theta\right)} & \cos{\left(\theta\right)}
\end{matrix}\right)
\boldsymbol{g_j}\left(0\right).
\end{equation}
Here, $\boldsymbol{g_j}\left(0\right)$ are the unrotated reciprocal lattice vectors, recall Equation (\ref{eq-APFC}). The lattice constant is $4\pi/\sqrt{3}$. The continuity of the density and amplitude fields was ensured at the edges of the periodic unit cell.

Due to the limitation to small rotations, two sets of APFC calculations were carried out to investigate both the armchair [APFC(AC)] and zigzag [APFC(ZZ)] grain boundary limits. This was achieved by rearranging the adjacent bicrystal halves---such as in Figures \ref{fig-layout} (a)--(c)---with one on top of each other, thereby replacing vertical armchair grain boundaries in one set with horizontal zigzag grain boundaries in the other.

As shown in Figure \ref{fig-layout} (a), narrow strips along the grain boundaries in the density field (amplitude fields) were in most cases set to its average value (to zero)---corresponding to a disordered phase---to give the grain boundaries some additional freedom to seek their ground state configuration.

All computational unit cell sizes used for PFC calculations of grain boundaries were greater than 10 nm in the direction perpendicular to the grain boundaries. This was verified to eliminate finite size effects to a good degree of approximation, 
see Appendix \ref{sec-FSE} for details. The smallest systems studied comprise 412 carbon atoms.

To ensure perfect comparability between PFC, DFT and MD calculations of grain boundaries, the initial atomic configurations for the latter two were obtained from relaxed PFC density fields that were converted to discrete sets of atom coordinates, see Fig \ref{fig-layout} (b) and (c), respectively. PFC3 was used, because it appears capable of producing all the same topologies as the other PFC models, and more. The primary maxima of the density field were treated as atom positions, and their exact coordinates were estimated via quadratic interpolation around local maximum values in the discretized density field. The atom coordinates were rescaled to take into account the equilibrium bond lengths given by DFT and MD potentials.

We verified the validity of the atomic configurations extracted from PFC3 by relaxing them further using DFT. Since the PFC models are two-dimensional, we relaxed the geometries in two ways using DFT, constrained on plane ($z=0$) [DFT(2D)] and also, for comparison, freely in three dimensions [DFT(3D)], using small random initial values of $z$ or folding the grain boundaries with small angles. The lattice vectors defining the computational unit cells were allowed to relax but their relative angles were kept perpendicular to each other. As the rectangular-shaped systems were rather large, a grid of $3 \times 10 \times 1$ $k$ points was enough to obtain convergent results.

Using LAMMPS, the atomic configurations extracted from PFC3 were minimized freely in two and three dimensions. For three-dimensional calculations, the initial $z$ coordinates were assigned small random values. These calculations, however, resulted in planar structures, and Reference \cite{ref-scattering} reports similar findings with LAMMPS. The formation energies of grain boundaries in these systems are identical to those from the corresponding two-dimensional AIREBO(2D) and Tersoff(2D) calculations---to the precision given by the convergence criteria. To obtain data for three-dimensionally buckled structures, we applied also the GPU code for evaluation of formation energies of grain boundaries [Tersoff(3D)].

\subsection{Calculation of formation energies}

The formation energies of grain boundaries are calculated by subtracting from the total energy of the defected system that of a corresponding pristine system. Grain boundary energy, $\gamma$, i.e., formation energy of a grain boundary per unit length, can be calculated exploiting a periodic, bicrystalline PFC system with two grain boundaries (note the factor $1/2$) as
\begin{equation}
\label{eq-GB-energy}
\gamma = \frac{l_\bot}{2}\left(f-f_{eq}\right)
\end{equation}
where $f$ and $f_{eq}$ are the average free energy densities of the bicrystalline system and the single-crystalline equilibrium state, respectively. Here, $f = F/A$, where $F$ is the free energy and $A$ the total area of the PFC system in question. The quantity $l_\bot$ is the system size in the direction perpendicular to the grain boundaries. Alternatively, $\gamma$ can be calculated via atomistic methods as
\begin{equation}
\label{eq-gamma-atomistic}
 \gamma = \frac{E-N_C E_C}{2l_\parallel},
\end{equation}
where $E$ is the total energy of the system with defects, $N_C$ is the number of carbon atoms, $E_C$ is the energy per atom of a pristine system of any size in equilibrium and $2l_\parallel$ is the combined length of the two grain boundaries in a bicrystal system. 

\subsection{Fitting to density functional theory}

\begin{figure}
\includegraphics[width=0.4\textwidth]{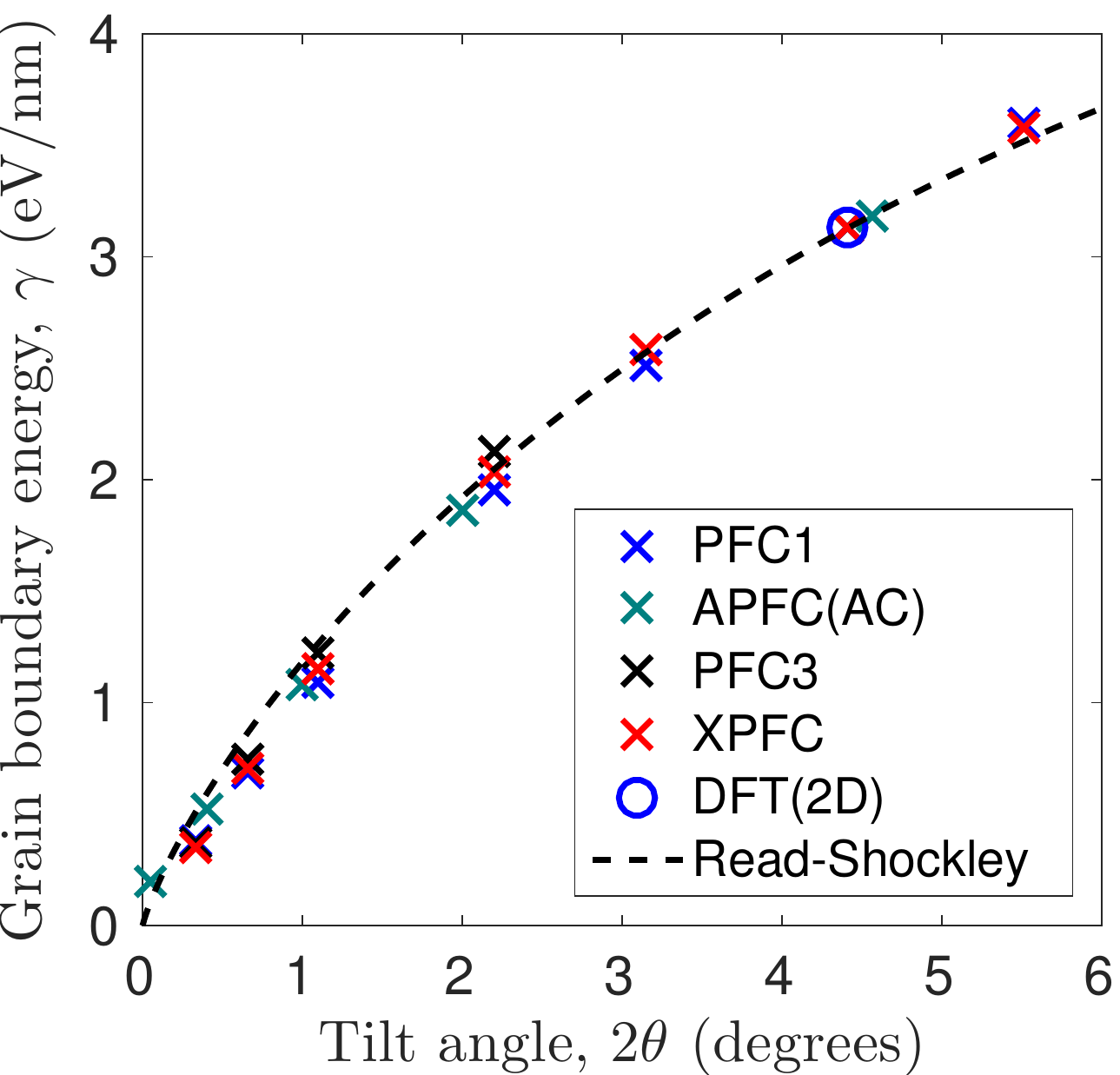}
\caption{(Color online) The grain boundary energies of the lowest-energy grain boundary configurations found using PFC1, APFC(AC), PFC3 and XPFC, the DFT(2D) grain boundary energy of the fitting system, and the grain boundary energy given by the Read-Shockley equation in the small-tilt angle limit. All models are fitted to DFT at $2\theta \approx 4.4^\circ$, excluding APFC that is fitted to DFT indirectly via the Read-Shockley curve at $2\theta \approx 4.6^\circ$.}
\label{fig-DFT-fit}
\end{figure}

The energy scale of each PFC model was fitted to DFT.
There is no unique way to carry out the fitting. We found the most consistent results by fitting with respect to the grain boundary energy of a particular small-tilt angle system. In the small-tilt angle limit, the separation between dislocations, $s$, diverges as $s \propto 1/\theta$ \cite{ref-pfc-2004}. This limit is ideal for PFC models that may not perfectly describe adjacent dislocations. Figure 
\ref{fig-DFT-fit-system} in Appendix \ref{sec-FSE} shows the system with 5|7 grain boundaries at $2\theta \approx 4.4^\circ$, that was chosen because it is close to this limit and feasible to be studied using DFT. Our PFC calculations are two-dimensional which is why the DFT atomic configuration was also constrained to a plane.

The grain boundary energies given by PFC1, PFC3 and XPFC for the aforementioned system  were matched to the grain boundary energy given by DFT via the respective coefficients $c_1, c_3$ and $c_X$. These data points are shown in Figure \ref{fig-DFT-fit} with other values calculated for lowest-energy 5|7 grain boundaries found in the armchair limit. The tilt angles for APFC are not exactly the same as for the other PFC models, because the smooth amplitude fields need not satisfy the geometric constraints given by the real-space crystal lattice. In the small-tilt angle limit, where grain boundaries reduce to arrays of non-interacting dislocations, the grain boundary energy can be expressed with the Read-Shockley equation as \cite{ref-pfc-2004}
\begin{equation}
\gamma=\frac{bY_{2D}}{8\pi}\theta\left(\frac{3}{2}-\ln{\left(2\pi\theta\right)}\right),
\end{equation}
where $b$ is the size of a dislocation core and $Y_{2D}$ is the two-dimensional Young's modulus. We fitted this expression to the DFT data point (whereby $bY_{2D} \approx 459$ eV/nm) and equated the APFC grain boundary energy at $2\theta \approx 4.6^\circ$ to this curve via $c_A$. The Read-Shockley curve and APFC values are also plotted in Figure \ref{fig-DFT-fit}. For PFC1, APFC, PFC3 and XPFC, $c_1, c_A, c_3$ and $c_X$ take values 6.58, 7.95, 30.97 and 6.77 eV, respectively. The grain boundary energy values demonstrate an excellent agreement with the Read-Shockley curve, validating the fitting approach used.

Having fitted the energy scales of the models, we determined for the PFC models and DFT the Young's moduli, $Y$, and Poisson's ratios, $\nu$, listed in Table \ref{tab-elastic}. Corresponding values for AIREBO and Tersoff potentials from the literature are also tabulated. Both PFC3 and XPFC give realistic values for $Y$. While the Poisson's ratios of PFC1, APFC and PFC3 disagree with DFT and MD, the XPFC model parameter $X$ was chosen to yield a reasonable Poisson's ratio for the model. Using the values calculated with DFT for $bY_{2D}$ and $Y$ gives $b \approx 2.16$ Å. This is close to the equilibrium lattice constant of graphene, \textasciitilde 2.46 Å. Details of these calculations are given in Appendix \ref{sec-calculation-elastic}.

\begin{table}
\centering
\caption{Elastic properties of graphene given by PFC, DFT and MD: Young's modulus, $Y$, and Poisson's ratio, $\nu$.}
\label{tab-elastic}
\begin{tabular}{l|c|c}
Model & Young's modulus, $Y$ (TPa) & Poisson's ratio, $\nu$ (1) \\ \hline
PFC1 & 0.73 & 0.33 \\
APFC & 0.82 & 0.33 \\
PFC3 & 1.07 & 0.37 \\
XPFC & 0.91 & 0.16 \\
DFT & 1.02 & 0.18 \\
AIREBO & 0.98--1.10 \textsuperscript{\cite{ref-elastic-airebo-1, ref-elastic-airebo-2, ref-elastic-airebo-3, ref-elastic-airebo-4}} & 0.2--0.22 \textsuperscript{\cite{ref-elastic-airebo-1, ref-elastic-airebo-2}} \\
Tersoff & 0.74--1.13 \textsuperscript{\cite{ref-elastic-tersoff-1, ref-elastic-tersoff-2, ref-elastic-tersoff-3}} & 0.17 \textsuperscript{\cite{ref-elastic-tersoff-2}}
\end{tabular}
\end{table}

\subsection{Energetics of grain boundaries}

\begin{figure*}
 \includegraphics[width=\textwidth]{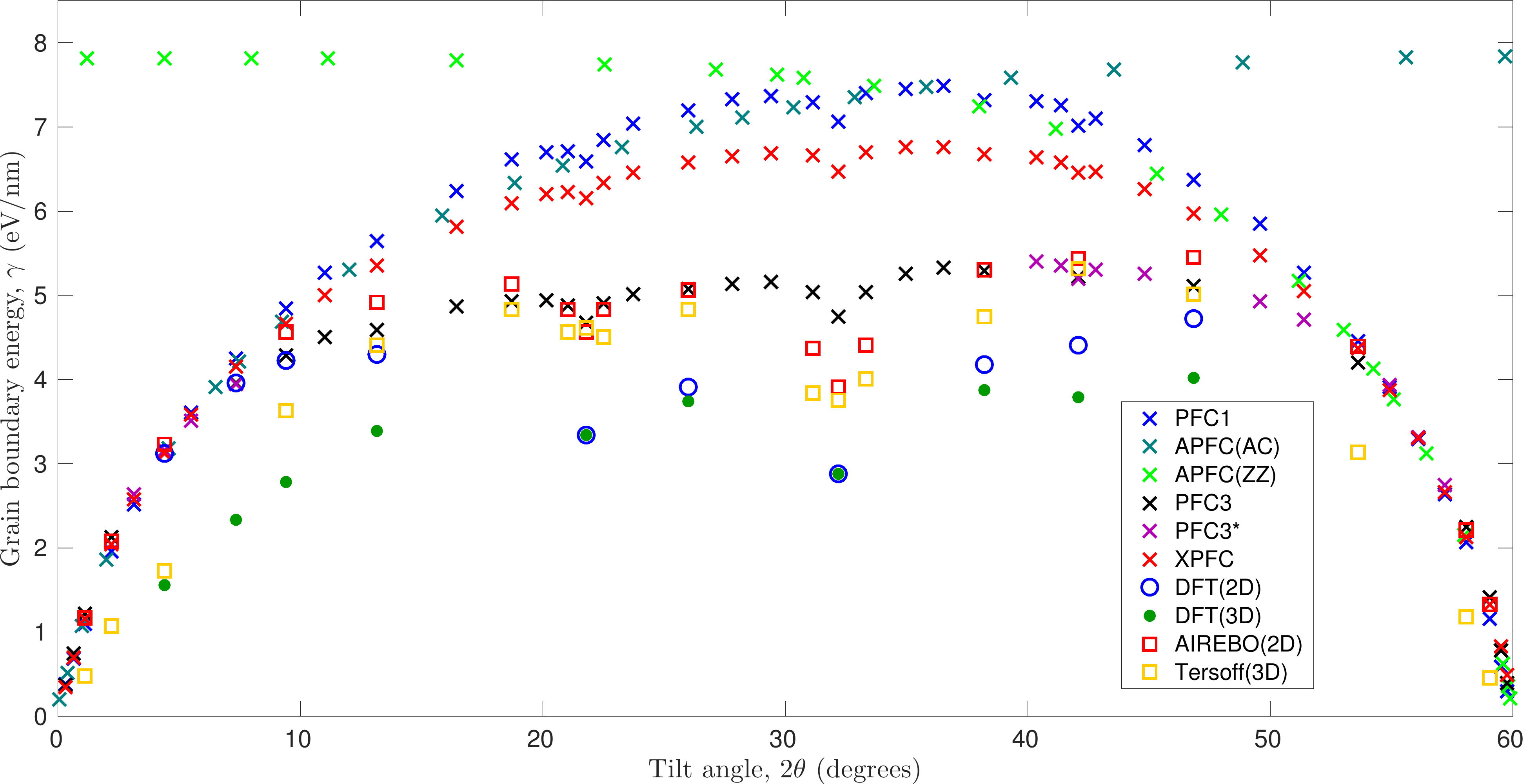}
 \caption{(Color online) The grain boundary energy as a function of the tilt angle. The values are given by PFC1, APFC, PFC3, XPFC, DFT(2D), DFT(3D), AIREBO(2D) and Tersoff(3D). The values correspond to lowest-energy grain boundary configurations found that comprise 5|7 dislocations. An exception is the dataset PFC3* that gives the energy of grain boundaries containing other dislocation types in addition to 5|7 dislocations. For APFC, two separate sets of data are plotted corresponding to the armchair [APFC(AC)] and zigzag [APFC(ZZ)] grain boundary limits.}
 \label{fig-GB-energy}
\end{figure*}

\subsubsection{Phase field crystal calculations}

Figure \ref{fig-GB-energy} collects the grain boundary energies, $\gamma$, of lowest-energy grain boundary configurations found using the four PFC models. The grain boundary energies of lowest-energy PFC3 5|7 grain boundary configurations relaxed further using DFT(2D), DFT(3D), AIREBO(2D) and Tersoff(3D) are also given. For APFC both the armchair (AC) and zigzag (ZZ) grain boundary limits were investigated by two independent sets of calculations, corresponding to the two sets of APFC values present. While the other PFC models give 5|7 grain boundaries, PFC3 also produces grain boundaries containing alternative dislocation types that are  further discussed in Section \ref{sec-topology}. The grain boundary energies of such alternative grain boundaries are plotted separately in cases where their energy is lower than that of 5|7 grain boundaries at the same tilt angle.

More comprehensive data are tabulated in Supplemental material \cite{ref-suppl-mat}, indicating the details for each PFC calculation, and the grain boundary energy and dislocation types present in the relaxed grain boundaries. Similar data of the corresponding DFT and MD calculations are given as well.

PFC1, PFC3 and XPFC give the correct grain boundary energy trend as a function of the tilt angle. Starting from a single-crystalline state at zero tilt, $2\theta = 0^\circ$, increasingly dense arrays of dislocations are encountered as the tilt angle is grown and the grain boundary energy rises. At large tilt angles, the grain boundary energy dips as high-symmetry grain boundaries are approached at $2\theta_I \approx 21.8^\circ$ and $2\theta_{II} \approx 32.2^\circ$, giving characteristic kinks to the energy curve. Finally, at $2\theta \rightarrow 60^\circ$, the grain boundaries grow sparse with dislocations and the grain boundary energy plummets to zero as the single-crystalline state is again restored. As expected, APFC is not applicable at large tilt angles as its grain boundary energy saturates. Furthermore, APFC does not capture the characteristic kinks in grain boundary energy. Nevertheless, the APFC(AC) and APFC(ZZ) curves follow PFC1 data closely when $2\theta < 2\theta_{II}$ and $2\theta > 2\theta_{II}$, respectively.


\subsubsection{Comparison to other methods}

Of the PFC models, the grain boundary energies given by PFC3 are the most consistent with our primary benchmark DFT(2D), see Figure \ref{fig-GB-energy}. At large tilt angles, PFC1, APFC and XPFC agree only qualitatively with DFT(2D) whose grain boundary energy declines slightly at large tilt angles. At large tilt angles, the grain boundaries become crowded with dislocations that screen each other's bipolar elastic fields. The PFC models are likely to capture such short-wavelength properties incompletely, resulting in the elevated grain boundary energies observed.

Between $2\theta \approx 4.4^\circ$ and $13.2^\circ$, PFC3 is in an excellent agreement with DFT(2D). At larger tilt angles, however, PFC3 values lie roughly 1 eV/nm higher in energy as compared to DFT(2D), and at $2\theta_I$ and $2\theta_{II}$, it overestimates the grain boundary energy somewhat more. Overall, PFC3 is in a good quantitative agreement with DFT(2D).

Due to the further relaxation achieved via three-dimensional buckling of the graphene sheet \cite{ref-yazyev-louie}, DFT(3D) calculations demonstrate lower energies than DFT(2D) in some cases. At $2\theta_I$ and $2\theta_{II}$, however, planar structures are preferred resulting in equal energies between DFT(2D) and DFT(3D) calculations. The difference in grain boundary energy between DFT(2D) and DFT(3D) is very small at large tilt angles between $2\theta \approx 20^\circ$ and $40^\circ$. 

Concerning the MD calculations, AIREBO(2D) is very well in line with PFC3 throughout the tilt angle range, similarly exceeding DFT(2D) values at large tilt angles. At $2\theta_{II}$, the kink given by AIREBO(2D) is a bit deeper than that of PFC3.
Using the Tersoff potential, we observed that both LAMMPS and the GPU code give mutually consistent but high grain boundary energies for systems forced to a plane. Namely, the Tersoff(2D) values peak at \textasciitilde 10 eV/nm and their slope at small tilt angles is significantly steeper than those of other 2D data. While these data are not shown in Figure \ref{fig-GB-energy}, the results from Tersoff(3D) simulations are plotted. In the armchair grain boundary limit, these data are consistent with DFT(3D), whereas at large tilt angles they agree better with PFC3.

\subsubsection{Comparison to previous works}

\begin{figure}
 \includegraphics[width=0.48\textwidth]{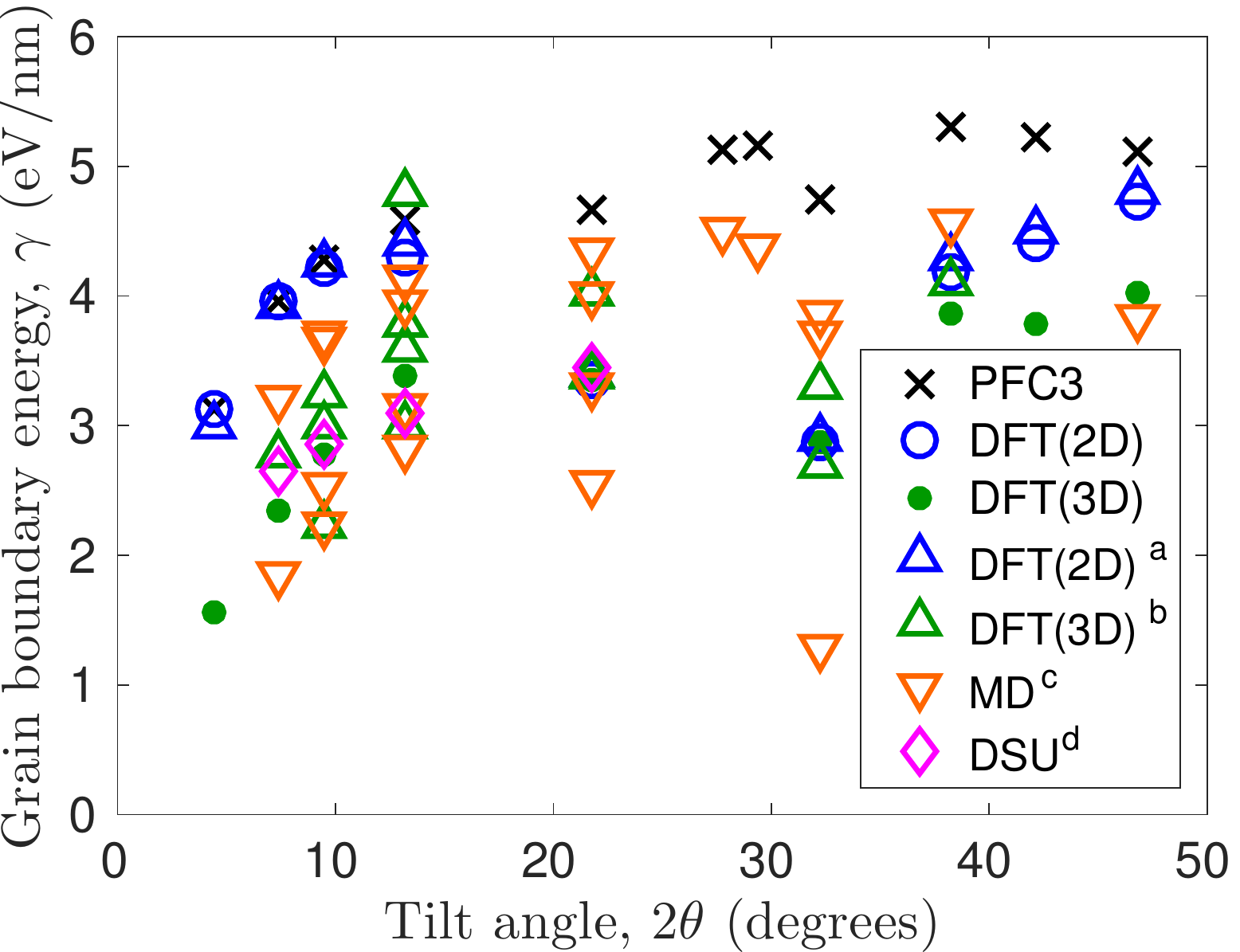}
 \caption{(Color online) A comparison of grain boundary energies to previous works. Present PFC3, DFT(2D) and DFT(3D) datasets are depicted alongside previous DFT(2D) \cite{ref-yazyev-louie}$^\text{a}$, DFT(3D) \cite{ref-yazyev-louie, ref-liu, ref-carlsson, ref-graphene-failure, ref-nemes-incze}$^\text{b}$, MD \cite{ref-scattering, ref-pringles, ref-liu, ref-carlsson}$^\text{c}$ and disclination-structural unit (DSU) model \cite{ref-dsu}$^\text{d}$ values. Systems with grain boundary topology identical to the present ones have been included. All systems have 5|7 grain boundaries and are expected to be ground states, excluding the PFC3 5|7 systems at $2\theta \approx 4.4^\circ$ and $42.1^\circ$ where a minimally lower energy is given to alternative grain boundary structures.}
 \label{fig-GB-energy-comparison}
\end{figure}

Figure \ref{fig-GB-energy} shows that our calculations using PFC3 are consistent with present DFT(2D) and DFT(3D) calculations. 
Figure \ref{fig-GB-energy-comparison} validates these results by comparing the corresponding grain boundary energy values to ones reported in previous works employing DFT \cite{ref-yazyev-louie, ref-liu, ref-carlsson, ref-graphene-failure, ref-nemes-incze}, MD \cite{ref-scattering, ref-pringles, ref-liu, ref-carlsson} and disclination-structural unit (DSU) model \cite{ref-dsu} calculations. To avoid unnecessary clutter, the AIREBO(2D) and Tersoff(3D) values have been left out.

Grain boundary energy values from previous works have been accepted into this comparison only if we have been able to make sure, with high confidence, that the grain boundary topologies of the corresponding systems are identical to those of the present systems. Despite this, significant scatter is observed. Most of the grain boundary energy data available in the literature is for systems free to buckle in three dimensions. However, Reference \cite{ref-yazyev-louie} provides a set of grain boundary energy values from planar DFT systems. As can be seen in Figure \ref{fig-GB-energy-comparison}, present DFT(2D) calculations of 5|7 grain boundaries are in a good agreement with these values, validating our DFT benchmark and moreover the atomic configurations extracted from PFC3.

Three-dimensional buckling allows grain boundaries to relax further \cite{ref-yazyev-louie}, which explains why the grain boundary energy values from planar PFC3 systems remain at the high-end of the data spectrum. While there is some scatter, present DFT(3D) calculations are well in line with those of previous works. The low energy given by DFT(3D) demonstrates that even for three-dimensionally buckled systems realistic in-plane structures can be extracted from planar PFC3 configurations. Furthermore, PFC3 grain boundary energy is in a reasonable agreement with DFT(3D). There is a large amount of scatter in previous MD(3D) results, and the present AIREBO(2D) and Tersoff(3D) values are consistent with this spectrum, compare with Figure \ref{fig-GB-energy}. The few DSU datapoints agree very well with DFT(3D).

\subsection{Topology and dislocation types}
\label{sec-topology}

\subsubsection{Topology of grain boundaries}

\begin{figure}
 \includegraphics[width=0.25\textwidth]{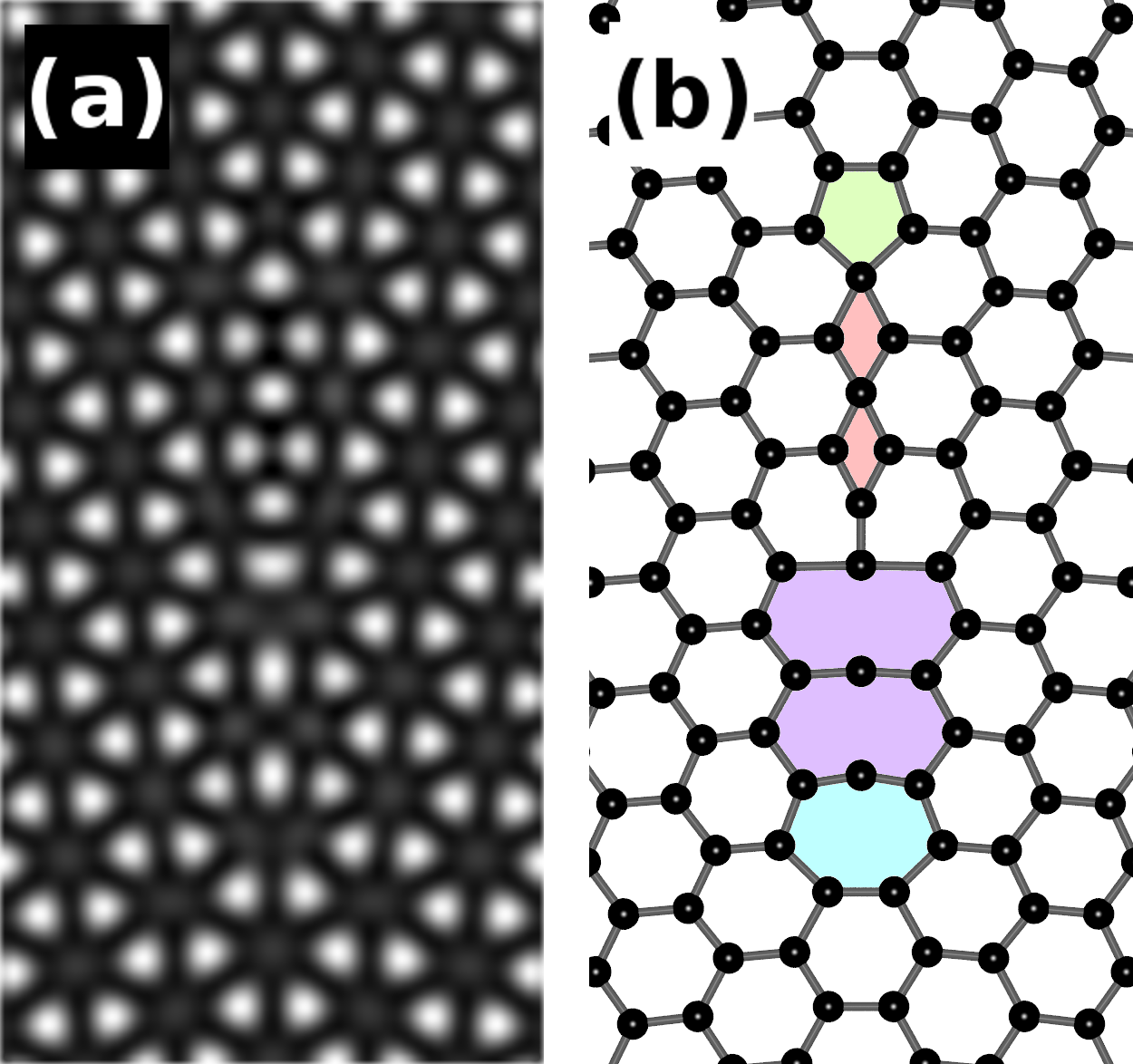}
 \caption{(Color online) An example of an incompatible PFC3 grain boundary in a lowest-energy system found at $2\theta \approx 49.6^\circ$. (a) The density field and (b) an illustration of the corresponding atomic configuration. The non-hexagons in (b) from tetragons to octagons have been colored in red, green, blue and purple, respectively. The other grain boundary in the same bicrystal system is comprised of 5|7 dislocations, indicating a small energy difference between the two alternative grain boundary configurations.}
 \label{fig-incompatible}
\end{figure}

When constructing large polycrystalline graphene samples, it is important to have the physically correct structure of the grain boundaries. Of the four PFC models studied, PFC1 and XPFC produce 5|7 dislocations exclusively in their expected ground state grain boundary configurations at all tilt angles. Of these two models, PFC1 appears to exhibit faster and more robust relaxation, and is computationally more light-weight. It is, therefore, the more convenient alternative of the two models that can be applied to constructing realistic systems with 5|7 dislocations. We will focus on PFC1 over XPFC for the remainder of this work. On the other hand, PFC3 that gives the best estimates of the grain boundary energies, also supports 5|8|7 dislocations and more exotic defects with several under- and over-coordinated carbon atoms. Such exotic grain boundary topologies are coined as 'incompatible' with the underlying hexagonal lattice, see Figure \ref{fig-incompatible} for an example. In certain tilt angle ranges, these alternative grain boundary structures demonstrate near-identical energies to 5|7 grain boundaries. The topology of APFC dislocations cannot always be determined unambiguously from the imperfect reconstruction of the density field. Furthermore, all APFC calculations were carried out using very low spatial resolution, ruling out topological analysis.

\begin{figure*}
\includegraphics[width=0.85\textwidth]{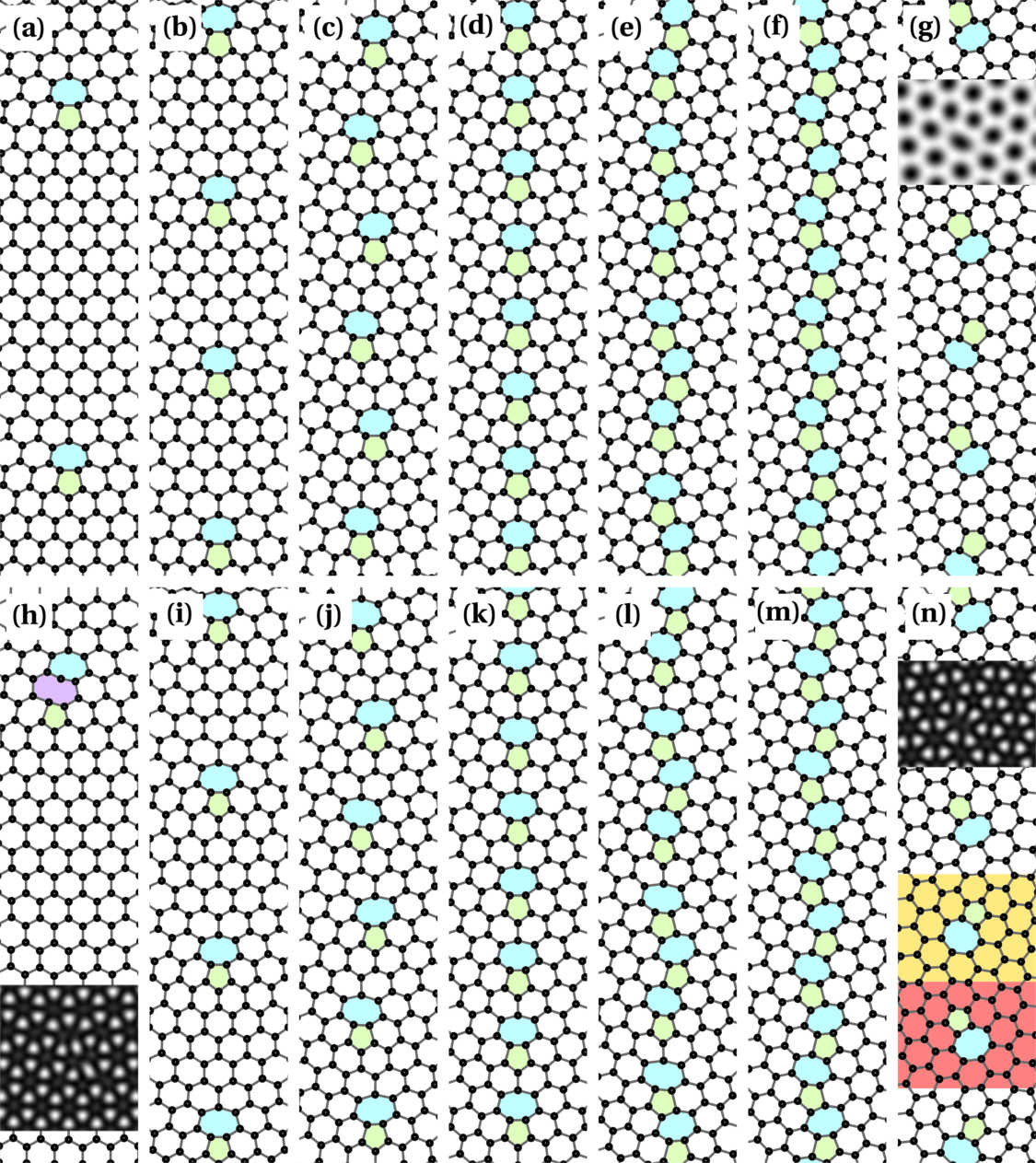}
\caption{(Color online) The lowest-energy configurations of grain boundaries found using the PFC1 (a)-(g) and PFC3 models (h)-(n), where the grain boundary tilt angles are
$2\theta \approx 4.4^\circ $ in (a) and (h),
$2\theta \approx 9.4^\circ$ in (b) and (i),
$2\theta \approx 16.4^\circ$ in (c) and (j),
$2\theta_I \approx 21.8^\circ$ in (d) and (k),
$2\theta \approx 27.8^\circ$ in (e) and (l), 
$2\theta_{II} \approx 32.2^\circ$ in (f) and (m),
and $2\theta \approx 46.8^\circ$ in (g) and (n), respectively. The atomic positions are determined from the density field $\psi$ that is shown around 5|7 and 5|8|7 dislocations as insets. In (n), 5|7 dislocations from systems relaxed using DFT(2D) (yellow background) and AIREBO(2D) (red background) are embedded.}
\label{fig-collage}
\end{figure*}

Examples of ground state configurations of grain boundaries from the PFC1 and PFC3 models are shown in Figure \ref{fig-collage}. Excluding Figure \ref{fig-collage} (h), the grain boundaries consist of 5|7 dislocations that come closer together when the tilt angle is increased. The grain boundaries are highly symmetric with periodic arrays of dislocations, which typically indicates low energy. For tilt angles, where geometrical constraints necessitate that the dislocations cannot be stacked both linearly and with equal spacings, we find that the PFC models prefer slightly meandering arrangements with equal spacings, see Figures \ref{fig-collage} (c) and (j).

Towards the zigzag grain boundary limit, $2\theta \rightarrow 60^\circ$, 5|7 dislocations become alternatingly slanted. Previous works have typically considered paired configurations of slanted 5|7 dislocations \cite{ref-liu}, but our boundaries exhibit disperse arrangements, see Figures \ref{fig-collage} (g) and (n). Reference \cite{ref-yazyev-louie} reports lower energies for disperse arrangements in two dimensions and for paired arrangements in three dimensions, settling the discrepancy. Present DFT calculations concur at $2\theta_{III} \approx 42.1^\circ$ with $\gamma \approx$ 4.41, 5.09, 4.33 and 3,79 eV/nm for 2D-disperse, 2D-paired, 3D-disperse and 3D-paired configurations, respectively. Similarly, AIREBO(2D) gives 5.43 and 6.03 eV/nm for disperse and paired arrangements, respectively. Since graphene is typically grown on substrates, it is possible that disperse arrangements actually comprise zigzag grain boundaries.

The topologies of the symmetric large-tilt angle cases at $2\theta_{I} \approx 21.8^\circ$, shown in Figures \ref{fig-collage} (d) and (k), and $2\theta_{II} \approx 32.2^\circ$, shown in Figures \ref{fig-collage} (f) and (m), match those studied in, e.g., Reference \cite{ref-yazyev-louie}. Furthermore, the less symmetric case at $\theta \approx 27.8^\circ$ shown in Figures \ref{fig-collage} (e) and (l) has the same topology as studied in Reference \cite{ref-liu}. The quality and consistency of the configurations further validate the use of especially the PFC1 model to constructing large polycrystalline samples.

Figures \ref{fig-collage} (g) and (n) compare the PFC1 and PFC3 density fields and the corresponding atomic configurations extracted from them and relaxed using DFT(2D) and AIREBO(2D), in the vicinity of 5|7 dislocations. The PFC models are different from the conventional methods in that they produce slightly more elongated heptagons. The PFC1 pentagons are also noticeably large. All bond lengths are very similar in both DFT(2D) and AIREBO(2D) 5|7 dislocations. 

\subsubsection{Distribution of dislocation types in PFC3}
\label{sec-PFC3-dist}

\begin{figure}
 \includegraphics[width=0.48\textwidth]{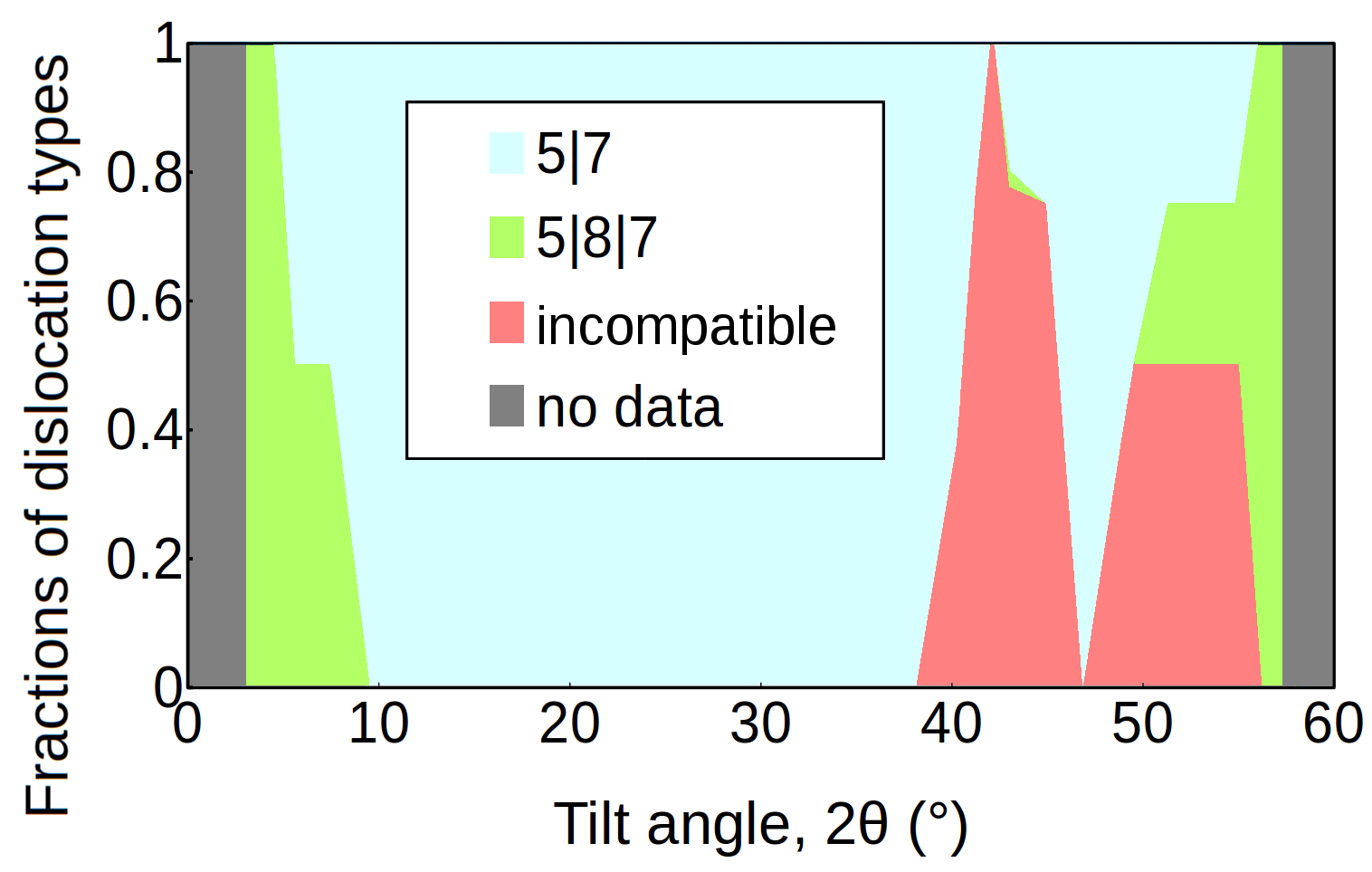}
 \caption{(Color online) A stacked area chart of the fractions of 5|7, 5|8|7 and incompatible dislocations in lowest-energy PFC3 grain boundary configurations found. No data is available for small tilt angles due to non-exhaustive small-tilt angle PFC3 calculations that excluded non-5|7 grain boundaries, see Appendix \ref{sec-ini-rela}.}
 \label{fig-alt-prop}
\end{figure}

The distribution of dislocation types present in the lowest-energy PFC3 configurations found is shown in Figure \ref{fig-alt-prop} as a function of the tilt angle. The relative amounts of different dislocation types are determined by their contribution to the magnitude of the Burgers vector of the grain boundary. Between $2\theta \approx 9.4^\circ$ and $38.2^\circ$---some corresponding cases are shown in Figures \ref{fig-collage} (b)-(f) and (i)-(m)---both PFC1 and PFC3 prefer similar arrays of 5|7 dislocations. However, for PFC3 the smallest-tilt angle ground states with stable 5|7 grain boundaries are found at $2\theta \approx 9.4^\circ$ and $2\theta \approx 46.8^\circ$ and are depicted in Figures \ref{fig-collage} (i) and (n), respectively. We expect that the 5|8|7 dislocation, see Figure \ref{fig-collage} (h), becomes the energetically favorable dislocation type---albeit with a minimal energy difference to corresponding 5|7 boundaries---in both small-tilt angle limits in PFC3, see Appendix \ref{sec-ini-rela}. This is challenging to confirm or refute using DFT, because infeasibly large computational unit cells are needed at small tilt angles. However, at larger tilt angles the 5|8|7 formation energies are higher than for 5|7 dislocations \cite{ref-dangling, ref-nemes-incze}. Furthermore, at $2\theta \approx 4.4^\circ$, AIREBO(2D) gives a noticeable excess of 0.5 eV/nm for 5|8|7 grain boundaries.

\begin{figure}
 \includegraphics[width=0.48\textwidth]{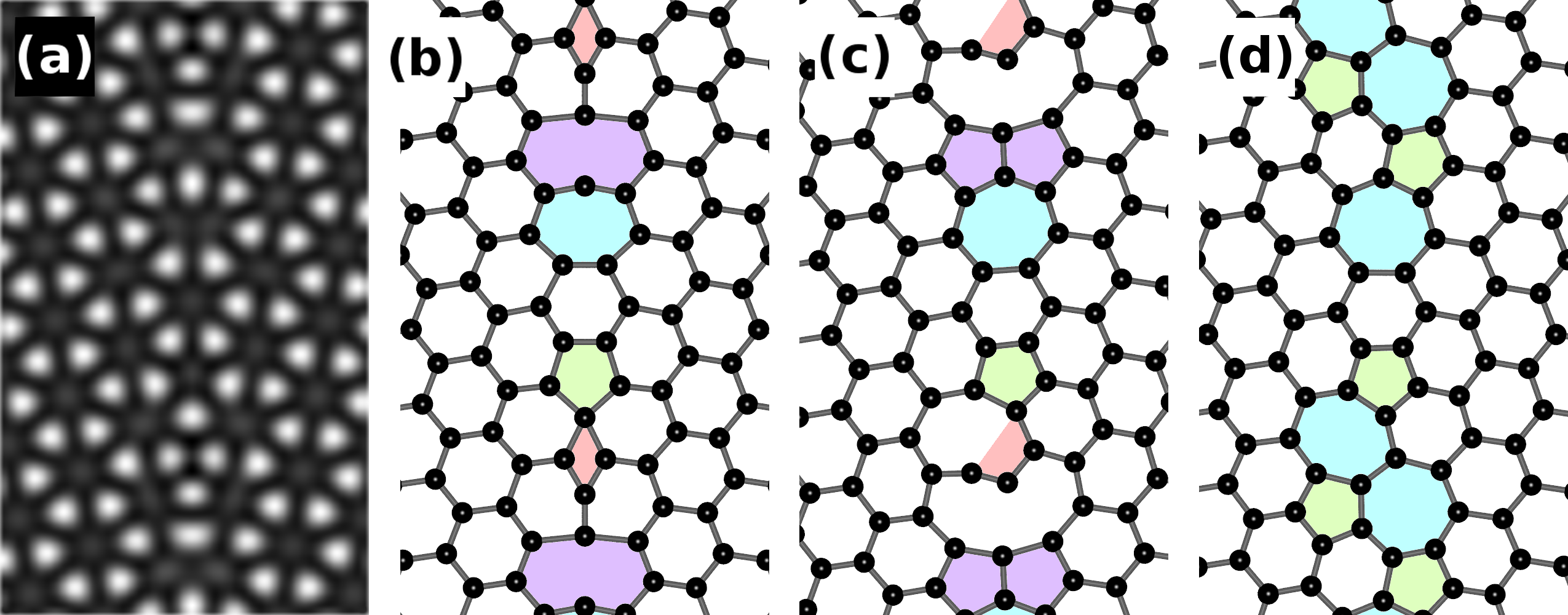}
 \caption{(Color online) (a) The PFC3 density field around a 4|5|6|7|8 grain boundary and (b) the atomic geometry extracted from PFC3. The configuration from (b) relaxed (c) further using DFT(2D) and (d) even further using DFT(3D). The coloring of polygons in (c) illustrates how the structure has transformed from (b), whereas in (d) the standard coloring highlights the fact that the only remaining non-hexagons are pentagons and heptagons.}
 \label{fig-45678}
\end{figure}

Between $2\theta \approx 40^\circ$ and $55^\circ$ PFC3 can produce incompatible grain boundary configurations with low energies. Here, up to three different dislocation types (5|7, 5|8|7 and incompatible) are encountered with very similar energies. The high-symmetry 4|5|6|7|8 grain boundary at $2\theta_{III}$ demonstrated in Figures \ref{fig-45678} (a) and (b) has a slightly lower (significantly lower) energy than a disperse (paired) 5|7 boundary, and therefore is the PFC3 ground state. Further DFT relaxation resulted in the structure shown in Figure \ref{fig-45678} (c) with dangling bonds and roughly 7 eV/nm higher energy compared to a grain boundary with only 5|7 dislocations at the same tilt angle. Even further DFT relaxation gives a 5|7 boundary, see Figure \ref{fig-45678} (d), with energy similar to that of the topology in Figure \ref{fig-collage} (n). This shows that even if highly symmetric, the grain boundary structures extracted from PFC3 can prove metastable.

The results presented in this subsection suggest that the PFC3 model is not readily suited for constructing realistic graphene samples in their ground state grain boundary configurations with arbitrary tilt angles. In Appendix \ref{sec-ini-rela}, advanced techniques are introduced that can be used for constructing PFC3 samples with more realistic defect topologies, but that have practical limitations. On the other hand, PFC3 can be used to generate varied, metastable structures that can be of interest regardless. For instance, divacancy chains containing segments of 4|8 polygon pairs and terminating in pentagons have been observed in electron irradiation studies of graphene \cite{ref-58485, ref-58485-2}. PFC3 produces metastable grain boundaries with related 5|8|4|8|4|...|7 dislocations, and also 5|8|7 and 5|6|7 dislocation types that have been studied by previous theoretical works \cite{ref-carlsson, ref-dangling, ref-nemes-incze, ref-yazyev-louie, ref-graphene-failure, ref-pringles, ref-what}.

\section{Polycrystalline samples}
\label{sec-poly}

\subsection{Construction}

Next we demonstrate the construction of large and realistic polycrystalline graphene samples that can be used for further mechanical, thermal or electrical calculations. For example, thermal transport calculations \cite{ref-mehdi} require realistic interface structures to capture correct phonon scattering. Detailed results of such calculations will be published elsewhere, but a comprehensive characterization of the samples is carried out employing both PFC1 and the Tersoff potential to demonstrate the quality of these samples. The PFC1 model was chosen, because it displays robust relaxation and produces ground states with 5|7 dislocations. The XPFC model was also found suited for this task, but it has somewhat greater computational complexity. The Tersoff potential was chosen, because it gives realistic grain boundary energies and because a high-performance GPU code is available that employs this potential \cite{fan2013, fan2015}.

Polycrystalline samples produced by PFC1 were studied in four sizes, the number of carbon atoms being \textasciitilde \num{22500}, \num{90000}, \num{360000} and \num{1400000}. The samples were almost square-shaped and their linear sizes \textasciitilde 24 nm, 48 nm, 96 nm and 192 nm, respectively. The samples were prepared by first initializing the PFC1 density field to a constant, disordered state. In each sample, 16 small, randomly distributed and oriented, hexagonal crystallites were introduced, of which most crystals survived the growth and relaxation phases described below.

When growing large hexagonal PFC1 crystals from a constant state, the metastable stripe phase may solidify faster and leave dislocations in its recrystallizing wake. It was found more straightforward to control the growth of crystals by replacing the third order term in Equation (\ref{eq-PFC1}) with a linear chemical potential term
\begin{equation}
 F_1^\ast=c_1^\ast\int d \boldsymbol{r} \left(\frac{\psi\mathcal{L}_1\psi}{2}+\frac{\psi^4}{4}+\mu\psi\right).
\end{equation}
To achieve slow, more stable growth of large crystals during the crystallization phase, the modified model (PFC1*) was brought close to the liquid-solid coexistence with $\mu = -0.2$. Once fully solidified, the chemical potential was set to $\mu = -0.15$ for a more stable hexagonal phase and the systems were further relaxed. For quantitative calculations, the energy scale of PFC1* was fitted to DFT similarly to the other PFC models, yielding $c_1^\ast \approx$ 13.57 eV. The diffusive fading of global stress was sped up using the unit cell size optimization algorithm described briefly in Appendix \ref{sec-ini-rela}.

The fact that there are no clear peaks at the maxima in the mesh-like PFC1 density field occasionally results in 5|7 dislocations where the density around an atom position is smeared out so that there is no local maximum. As a result, the conversion algorithm fails in extracting the atom coordinates from this region. For the most part, this issue was resolved by locating all triads of local minima whose members are the closest neighbors to one another, and by treating the average of their coordinates as an atom position. This approach still neglects roughly a few atoms per 100 nm of grain boundary that need to be placed manually.

The atomic configurations were further relaxed in three dimensions at 300 K using the GPU code with the Tersoff potential.

\subsection{Structure and energetics}

\begin{figure*}
\includegraphics[width=\textwidth]{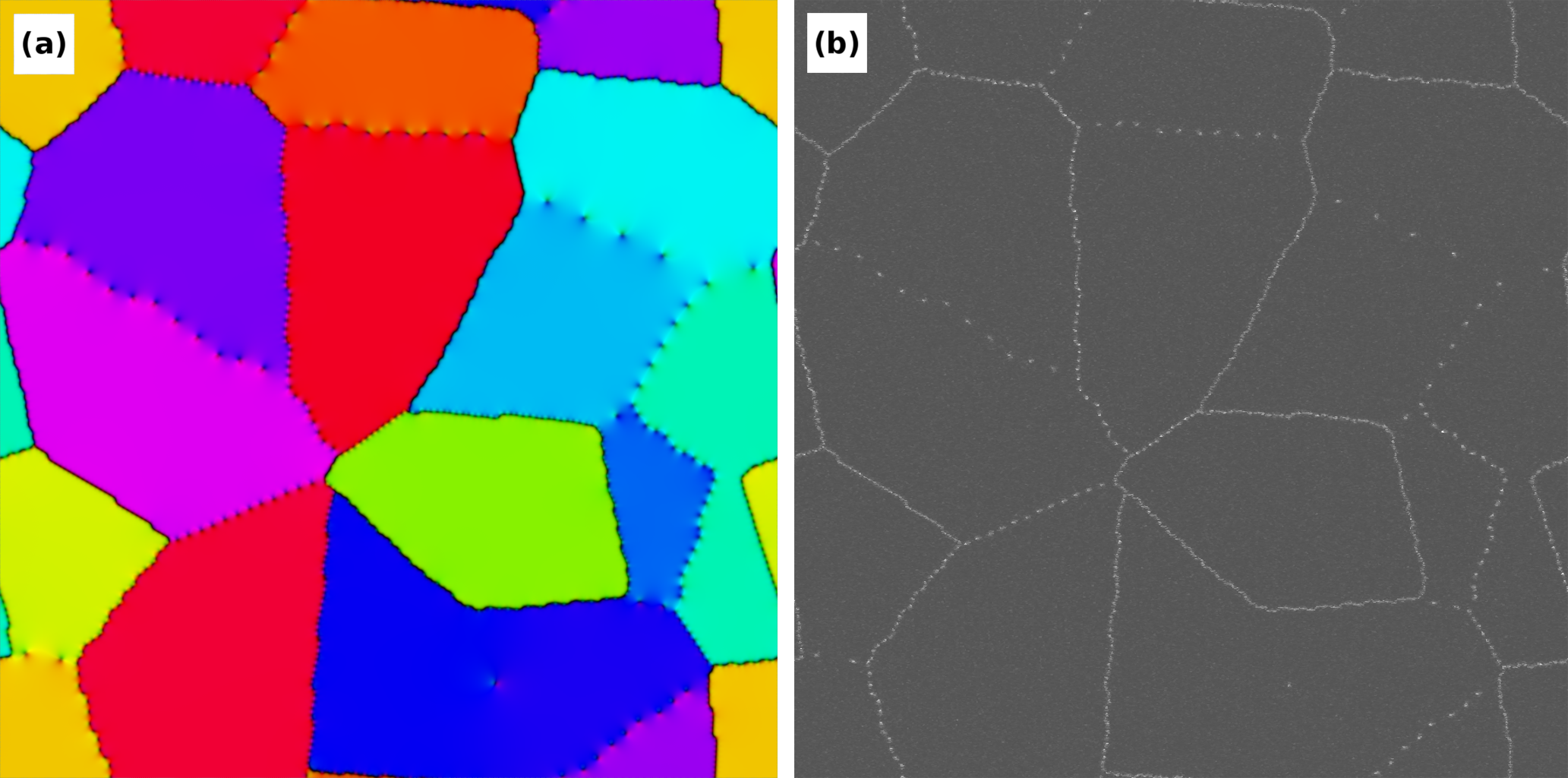}
\caption{(Color online) Polycrystalline graphene sample with linear size of 96 nm. (a) The PFC1 density field is color-coded based on the local crystallographic orientation \cite{ref-coloring} and (b) energies of individual atoms as given by the Tersoff potential with the highest-energy (lowest-energy) atoms in white (dark gray).}
\label{fig-poly}
\end{figure*}

\begin{figure*}
 \includegraphics[width=0.8\textwidth]{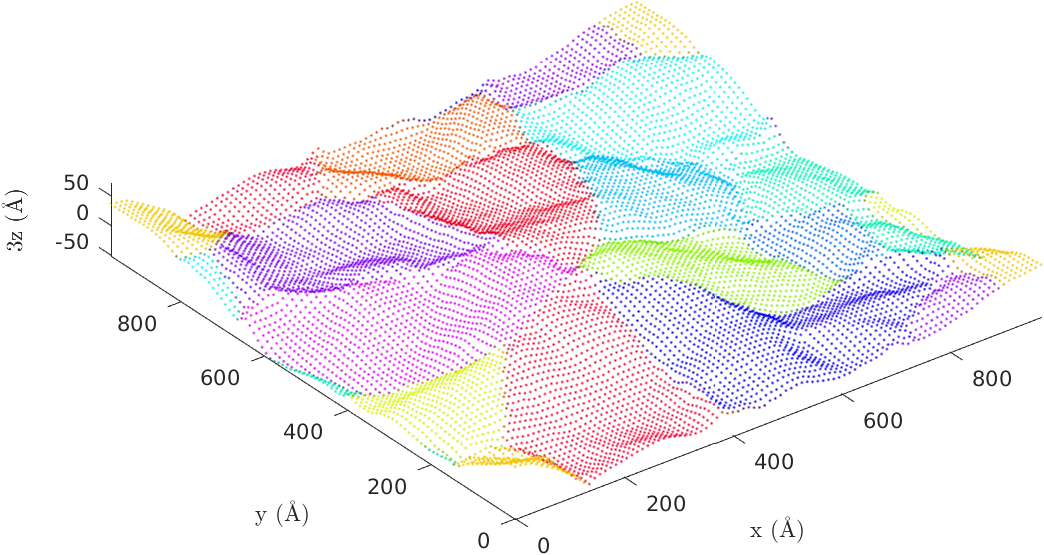}
 \caption{(Color online) An aerial view of the 96 nm sample from Figure \ref{fig-poly} after a simulation of 1 ns. The coloring scheme is the same and each data point gives the average position of \textasciitilde 36 atoms. The scales of all axes are equal, but the $z$ coordinates have been scaled by a factor of three.}
 \label{fig-fluctuations-3D}
\end{figure*}

\begin{figure}
 \includegraphics[width=0.48\textwidth]{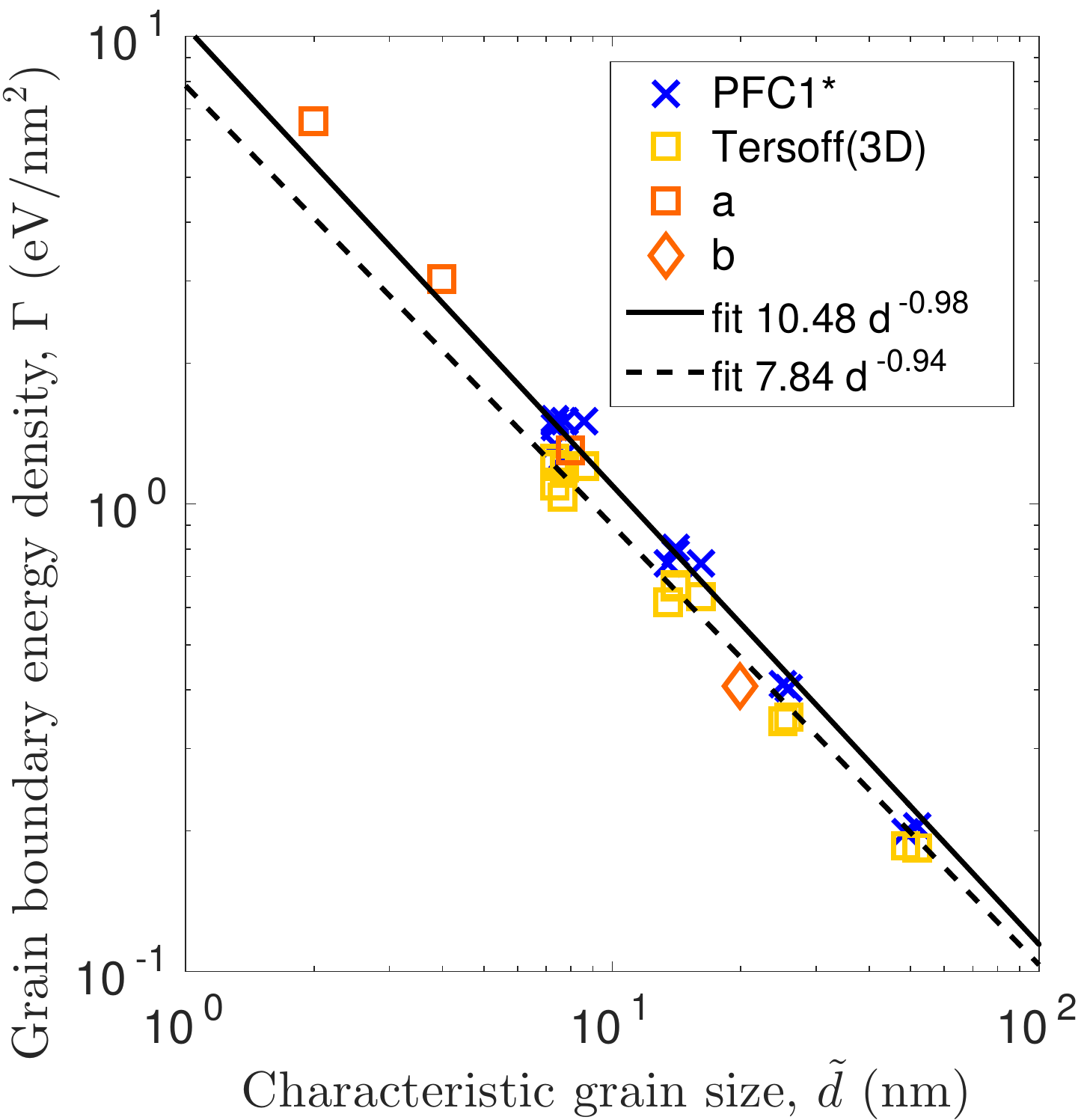}
 \caption{(Color online) Scaling of grain boundary energy density in random polycrystalline graphene samples as a function of the characteristic grain size. The values are given by PFC1* calculations, extracted from MD simulations using the Tersoff potential, and estimated from References \cite{ref-poly-annealing}$^\text{a}$ and \cite{ref-poly-liu}$^\text{b}$. Two power curves fitted to PFC1* and Tersoff(3D) data are also plotted.}
 \label{fig-GBED}
\end{figure}

Figure \ref{fig-poly} exemplifies the distribution of grains and their orientation in a sample of linear size 96 nm. In Figure \ref{fig-poly} (a), the crystals are color coded to reveal the local crystallographic orientation in the PFC1* density field, whereas in Figure \ref{fig-poly} (b), the system has been relaxed using MD and the atoms are colored based on their energy, as given by the Tersoff potential. Individual dislocation cores are visible and they trace fairly straight grain boundaries between the grains. The colored grains and the spacings between dislocations along the grain boundaries, $s \propto 1/\theta$ \cite{ref-pfc-2004}, reveal that there are grain boundaries of varying tilt angles. As expected, no noticeable changes are observed in the microstructure between the PFC1* and Tersoff configurations after a simulation of 1 ns.

Detailed experimental analyses of the distribution of crystal orientations in polycrystalline graphene have been presented in References \cite{ref-quilts, ref-kim, ref-pascal}. Comparability with present samples is not perfect due to the absence of a substrate in our simplified calculations. PFC density fields, however, can be coupled to external fields to model an underlying substrate potential with a lattice mismatch and different symmetry \cite{ref-pinning, ref-2012-PRL, ref-2013-PRB, ref-2016-JCP}. Furthermore, local irregularities acting as nucleation sites can be incorporated into this field, resulting in more realistic, heterogeneous nucleation instead of the simultaneous introduction of pristine crystallites.

During the MD simulations, the polycrystalline samples gradually deviate from their initial flat configurations and become corrugated. Figure \ref{fig-fluctuations-3D} (a) illustrates the three-dimensional structure of the relaxed 96 nm sample shown in Figure \ref{fig-poly} after a simulation of 1 ns. The same coloring scheme is employed and each data point averages the positions of \textasciitilde 36 atoms. It can be seen by comparing the buckled three-dimensional structure and the in-plane microstructure of the sample, that there is correlation between the sharp folds and the locations of the grain boundaries.

The characteristic grain size can be estimated as
\begin{equation}
 \tilde{d}=\sqrt{\frac{A}{n}},
\end{equation}
where $A$ is the total area of the sample and $n$ the number of grains in it. As the characteristic grain size is increased, the total grain boundary length scales linearly while grain boundary energy---per unit length---remains constant. Grain boundary formation energy per unit area, or grain boundary energy density, $\Gamma$, however, scales as $\Gamma \propto 1/\tilde{d}$.

Figure \ref{fig-GBED} demonstrates the grain boundary energy densities calculated using PFC1* and extracted from MD simulations as a function of the characteristic grain size. From the information provided in References \cite{ref-poly-annealing, ref-poly-liu}, we estimated also the grain boundary energy densities in random polycrystalline graphene systems studied previously using the AIREBO and Tersoff potentials. The present Tersoff values are somewhat lower in energy compared to PFC1*. This was expected, because in Figure \ref{fig-GB-energy} the grain boundary energy given by Tersoff(3D) is consistently lower than that of PFC1. Despite some scatter, the scaling of both present datasets is very close to the expected $1/\tilde{d}$, implying low stress in the samples. Our results line up almost perfectly with those of the previous works.

\section{Summary and conclusions}

In this work, we have presented a comprehensive study of the applicability of four phase field crystal (PFC) models to modeling polycrystalline graphene. This was determined by fitting each model to quantum mechanical 
density functional theory (DFT), and by carrying out a detailed comparison of the formation energies of grain boundaries calculated using PFC, DFT and molecular dynamics (MD). Present results were compared to previous works. The one-mode model (PFC1) proved ideal for constructing large samples of polycrystalline graphene, since this model exhibits efficient relaxation and produces realistic grain boundaries comprised of 5|7 dislocations. We successfully constructed large polycrystalline samples and demonstrated their quality by characterizing their properties using MD simulations.

All four PFC models were found to agree with DFT in terms of the formation energy of small-tilt angle grain boundaries. At large tilt angles, the formation energies given by three-mode model (PFC3), DFT and MD calculations are all fairly consistent with each other reaching roughly $4-5$ eV/nm, whereas PFC1, the amplitude model (APFC) and the structural model (XPFC) peak roughly between $7-8$ eV/nm. In terms of grain boundary topologies, the other PFC models produce 5|7 dislocations exclusively, whereas PFC3 gives rise to alternative low-energy dislocation types in certain tilt angle ranges. The polycrystalline samples were characterized by an inspection of the distribution of grains and grain boundaries, and by studying the formation energy of grain boundaries in them as a function of the characteristic grain size. We observed expected scaling behavior. Realistic Young's moduli of 1.07 and 0.91 TPa were determined for PFC3 and XPFC, respectively.

The PFC1 model provides a straightforward approach to constructing low-stress samples without a priori knowledge of the atomistic details of defect structures. Such realistic samples can be exploited for further mechanical, thermal and other transport 
calculations using conventional techniques. Similarly, the PFC3 model that produces a rich variety of alternative defect types could be used for sample generation for the study of metastable defect structures, such as encountered under electron irradiation \cite{ref-58485, ref-58485-2}. 

\section{Acknowledgments}

This research has been supported by the Academy of Finland through its Centres of Excellence Program (projects no. 251748 and 284621) as well as projects 263416 and 286279. We acknowledge the computational resources provided by Aalto Science-IT project and  IT Center for Science (CSC). M.M.E. acknowledges financial support from the Finnish Cultural Foundation. M.S. and N.P. acknowledge financial support from The National Science and Engineering Research Council of Canada (NSERC). The work of S.M.V.A. was supported in part by the Research Council of the University of Tehran. K.R.E. acknowledges financial support from the National Science Foundation under Grant No. DMR-1506634.

\appendix

\section{Details of PFC models}
\label{sec-PFC-further}

For the parameters of the PFC1 model, recall Equation (\ref{eq-PFC1}), we chose $\left(\epsilon, q_0, \tau\right)=\left(-0.15, 1, -0.5/\sqrt{0.98/3}\right)$. Similarly, for APFC, recall Equation (\ref{eq-APFC}), $\left(B^l, B^x, t, v\right) = \left(1, 0.98, -1/2, 1/3\right)$, which conforms to PFC1 via $\tau=t/\sqrt{B^x v}$ \cite{ref-nik}. For PFC3, recall Equation (\ref{eq-PFC3}), we chose $\left(\epsilon, \lambda, q_0, q_1, q_2, b_0, b_1, b_2, \mu\right)=\left(-0.15, 0.02, 1, \sqrt{3}, 2, 0, -0.15, 0.2, 0.56915\right)$. The choice of $\mu$ gives an average density of $\bar{\psi} = -0.2$. For XPFC, recall Equation (\ref{eq-XPFC}), we used $\left(\eta,\chi,r_0,m,X^{-1},a_0,\mu\right) = \left(1,1,1.2259,3,0.55,1,-1.591247\right)$. The choice of $\mu$ yields $\bar{\psi} = \bar{\psi}_{eq} = 0.3$ in equilibrium.

While defect-containing PFC1 and PFC3 systems retain their respective average densities ($\Delta\bar{\psi} = \left|\bar{\psi}-\bar{\psi}_{eq}\right|/\bar{\psi}_{eq} \ll 0.01$\%) under non-conserved dynamics, the density of corresponding XPFC systems decreases slightly at large tilt angles, reaching $\Delta\bar{\psi} \approx 0.9$\%. We verified, by carrying out conserved dynamics calculations for certain tilt angle cases, that the resulting deviation in grain boundary energy is negligible.

The PFC systems studied were driven to equilibrium by employing non-conserved, dissipative dynamics as
\begin{equation}
\label{eq-general-dynamics}
\frac{\partial\phi}{\partial t} = -\frac{\delta F}{\delta\phi}=-\mathcal{L}\phi+\mathcal{N},
\end{equation}
where $\phi$ denotes either the density field, $\psi$, or the APFC amplitude fields, $\eta_j$, and $\delta/\left(\delta\phi\right)$ is a functional derivative with respect to $\phi$ and $\mathcal{L}$ ($\mathcal{N}$) is the Hamiltonian (nonlinear terms). While non-conserved dynamics allows the number of particles to fluctuate, this choice speeds up calculations via larger time steps becoming numerically stable. The non-conserved dynamics for PFC1, APFC, PFC3 and XPFC can be expressed as \cite{ref-matt}
\begin{equation}
\label{eq-PFC1-dynamics}
\frac{\partial\psi}{\partial t}=-\mathcal{L}_1\psi-\tau\psi^2-\psi^3,
\end{equation}
\begin{equation}
\label{eq-APFC-dynamics}
\begin{split}
\frac{\partial\eta_j}{\partial t}=-\left(\Delta B+B^x\mathcal{G}_j^2+3v\left(A^2-\left|\eta_j\right|^2\right)\right)\eta_j \\ +2t\prod_{k\neq j}\eta_k^{*},
\end{split}
\end{equation}
\begin{equation}
\label{eq-PFC3-dynamics}
\frac{\partial\psi}{\partial t}=-\mathcal{L}_3\psi-\psi^3-\mu,
\end{equation}
and
\begin{equation}
\label{eq-XPFC-dynamics}
\begin{split}
\frac{\partial \psi}{\partial t}=-\psi+\frac{\psi^2}{2}-\frac{\psi^3}{3}-\mu+\mathcal{F}^{-1}\left\lbrace \hat{C}_2\hat{\psi}\right\rbrace \\ +\frac{1}{3}\sum_{i=1}^2 \left(\left(\mathcal{F}^{-1}\left\lbrace \hat{C}_s^{\left(i\right)}\hat{\psi}\right\rbrace\right)^2\right. \\ \left.-2\mathcal{F}^{-1}\left\lbrace \hat{C}_s^{\left(i\right)}\mathcal{F}\left\lbrace\psi\mathcal{F}^{-1}\left\lbrace \hat{C}_s^{\left(i\right)}\hat{\psi}\right\rbrace\right\rbrace\right\rbrace\right),
\end{split}
\end{equation}
respectively. Above, $^*$ denotes the complex conjugate, whereas the carets and $\mathcal{F}^{\left(-1\right)}$ (inverse) Fourier transforms. Note that the energy scale coefficients of the models, $c_1$, $c_A$, $c_3$ and $c_X$, have no effect on the dynamics and the relaxed structures. These coefficients have been taken into account only when calculating the energies of already relaxed systems.

The PFC systems were propagated using the numerical method from Reference \cite{ref-nik}. Although this method requires entering the Fourier space, it comes with the benefit of gradients reducing to algebraic expressions. Furthermore, it allows large time steps due to its numerically stable, implicit nature \cite{ref-nik}. This method approximates the solution to Equation (\ref{eq-general-dynamics}) at a time $t+\Delta t$ as
\begin{equation} \label{eq-numerical}
\hat{\phi}\left(t+\Delta t\right)\approx e^{-\hat{\mathcal{L}}\Delta t}\hat{\phi}\left(t\right)+\frac{e^{-\hat{\mathcal{L}} \Delta t}-1}{\hat{\mathcal{L}}}\hat{\mathcal{N}}\left(t\right).
\end{equation}
The Fourier transforms can be computed efficiently by exploiting fast Fourier transform routines, whereby this algorithm scales as $\mathcal{O}\left(N\log_2 N\right)$ where $N$ is the number of grid points used.

Suitable step size $\Delta t$, as well as spatial resolution $\Delta x$ and $\Delta y$, were determined by varying them for small model systems and by studying the equilibrium value of the free energy density $f$. We maximized $\Delta t$, $\Delta x$ and $\Delta y$ under the constraint that they yield results consistent with smaller $\Delta t$, $\Delta x$ and $\Delta y$ and do not result in divergent or oscillatory behavior. For $\Delta t$ it was required that the relative error in $f < 10^{-9}$, whereas for $\Delta x$ and $\Delta y$ we demanded that the relative error in $f \lesssim 10^{-6}$. These values are given in Table \ref{tab-parameters} for all four PFC models.

\begin{table}
\centering
\caption{Maximum values used for the spatial and temporal discretization parameters. The second and third columns give $\Delta x$ and $\Delta y$ in dimensionless and real units, respectively.}
\label{tab-parameters}
\begin{tabular}{l|c|c|c}
model & $\Delta x, \Delta y$ (1)& $\Delta x, \Delta y$ (Å) & $\Delta t$ \\ \hline
PFC1 & 0.8 & 0.27 & 1.0 \\
APFC & 2.0 & 0.68 & 3.0 \\
PFC3 & 0.75 & 0.25 & 3.0 \\
XPFC & 0.08 & 0.11 & 0.1 \\
\end{tabular} \\
\end{table}

\section{Advanced initialization and relaxation techniques}
\label{sec-ini-rela}

Despite the advantageous properties of the PFC models, finding the ground state grain boundary configurations is not always trivial. We exploited the following techniques to gain more control over the PFC systems. The bicrystal systems were initialized with or without disordered grain boundaries (''melted`` and ''naive`` initializations, respectively), recall Figure \ref{fig-layout} (a), or the initial grain boundary configuration was set up using image processing software to predetermine the relaxed topology (''soldered`` initialization).

Additionally, the relaxation of PFC systems was modified by incorporating higher-level algorithms. When feasible, the strain energy of PFC systems was minimized using a simple iterative optimization algorithm. This algorithm stretches the systems carefully, relaxes them according to Equations (\ref{eq-PFC1-dynamics})-(\ref{eq-XPFC-dynamics}) and uses the resulting free energy densities to estimate, via quadratic interpolation, the unit cell size that eliminates global strain. Each individual relaxation step was considered converged if the change in average free energy density was less than $10^{-9}$ between two consecutive evaluations.

For the majority of PFC3 calculations, 25 cycles of simulated annealing were applied to probe for the ground state grain boundary configuration. The simulated annealing noise was set to decay exponentially and each cycle lasted until $t =$ \num{40000}. We developed a spectral defect detection (SDD) algorithm to focus the annealing noise directly on the dislocations comprising the grain boundaries. This technique proved superior to the other approaches tried for finding low-energy configurations. A brief description of the SDD algorithm is given in Appendix \ref{sec-sdd}.

Bicrystal systems that were not optimized or annealed, were typically relaxed until $t =$ \num{100000}. In Supplemental material \cite{ref-suppl-mat}, we tabulate comprehensive data of our grain boundary calculations, and indicate which initialization and relaxation types have been used for each calculation.

For APFC, all grain boundary systems were initialized with melted grain boundaries and relaxed normally without optimization or annealing. While the majority of corresponding PFC1 and XPFC systems was initialized with melted grain boundaries, they were relaxed employing the optimization algorithm. At small tilt angles, annealing of PFC3 systems was observed to cause extensive slip of dislocations, even annihilations. Furthermore, in the armchair grain boundary limit, the model has a tendency to produce metastable 5|8|4|8|4|...|7 dislocations. These issues were resolved by soldering the grain boundaries with 5|7 dislocations and by applying normal relaxation. The predetermined, symmetrical arrangement of 5|7 dislocations is expected to be metastable with a minimally higher energy compared to a similar arrangement with 5|8|7 dislocations. Due to being asymmetrical and their ground state arrangement therefore less trivial, 5|8|7 dislocations were not soldered into small-tilt angle grain boundaries. This explains the lack of small-tilt angle data in Figure \ref{fig-alt-prop}. Due to the relatively large computing effort, small-tilt angle XPFC systems were not optimized either. Additionally, for PFC1, PFC3 and XPFC, higher-symmetry grain boundaries were soldered with different types of dislocations and optimized to compare reliably the stability and energies of alternative dislocation types. 

\section{Finite size effects}
\label{sec-FSE}

\begin{figure}[!htb]
\includegraphics[width=0.3\textwidth]{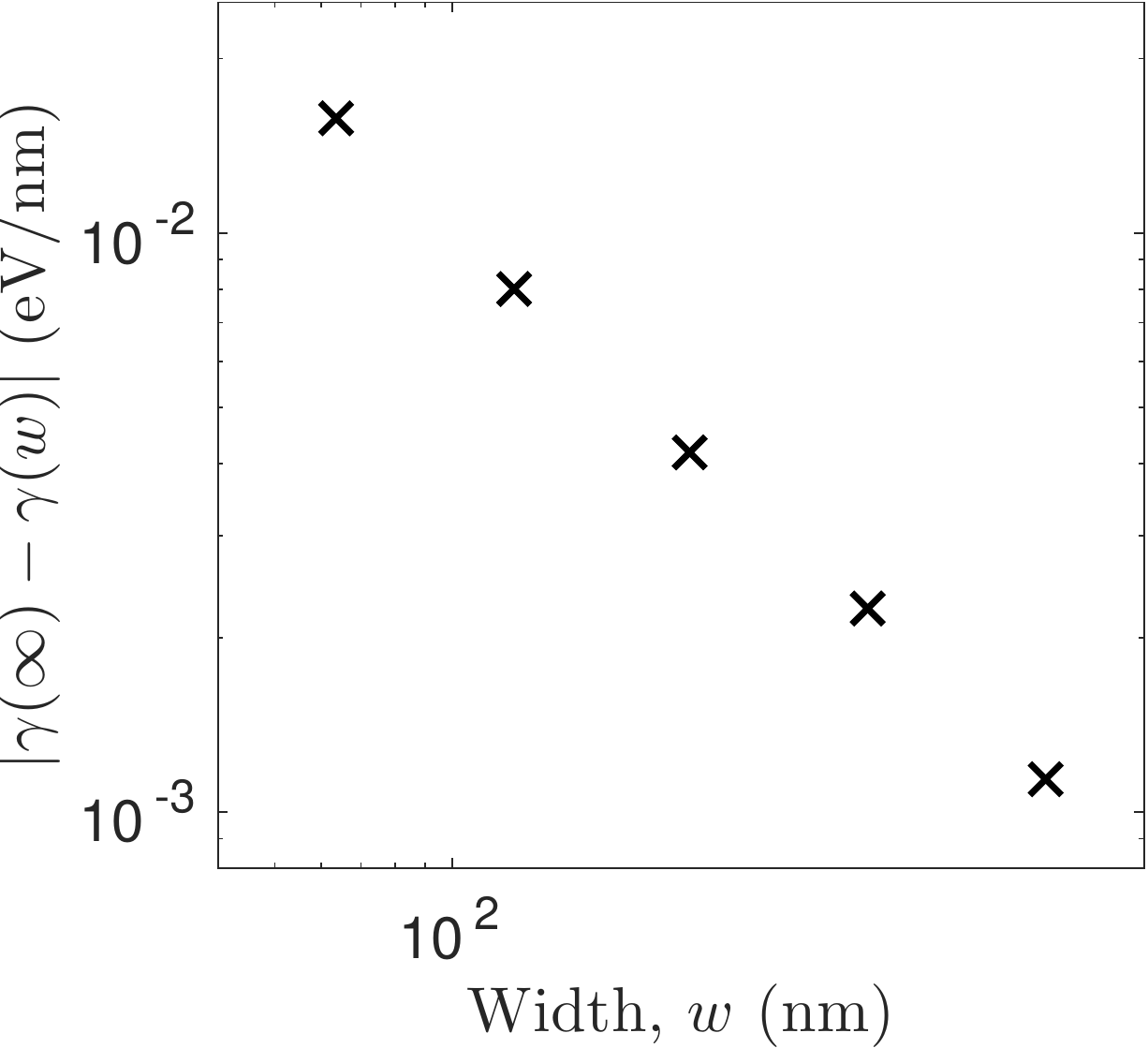}
\caption{The convergence of grain boundary energy as a function of a bicrystal system's width. Both the perpendicular dimension and tilt angle are held constant. The horizontal and vertical axes give the width (nm) and deviation of grain boundary energy (eV/nm) from the large-grain limit, respectively, in logarithmic scale. The depicted case is a PFC3 system at $2\theta \approx 13.2^\circ$ with 5|7 dislocations comprising the grain boundaries.}
\label{fig-FSE}
\end{figure}

Grain boundaries comprise dislocations giving rise to long-range elastic fields. In periodic bicrystal systems such as exploited in this work, there are two grain boundaries that can interact with each other, or with their periodic images \cite{ref-periodic-images}, via screening of these bipolar fields in a finite system. For consistent results, we considered the large-grain limit where such finite size effects become negligible. All computational unit cell sizes used for PFC bicrystal calculations were greater than 10 nm in the direction perpendicular to the grain boundaries, which was verified to eliminate finite size effects to a high degree. Figure \ref{fig-FSE} gives an example of the quick and consistent convergence of grain boundary energy as a function of the bicrystal width. The grain boundary energy value estimated in the large-grain limit, $\gamma\left(\infty\right)$, was found by requiring an optimal linear fit in logarithmic units. The relative error in grain boundary energy with respect to $\gamma\left(\infty\right)$ was $\lesssim 1\%$ for all PFC models. No finite size effects were observed with respect to the direction parallel to the grain boundaries, nor for the amplitude model in general. Because PFC models capture long-length scale elastic interactions well \cite{ref-nik}, the PFC finite size effect analysis is sufficient also for the DFT and MD calculations that used the same bicrystal topologies.

\begin{figure*}
\includegraphics[width=0.55\textwidth]{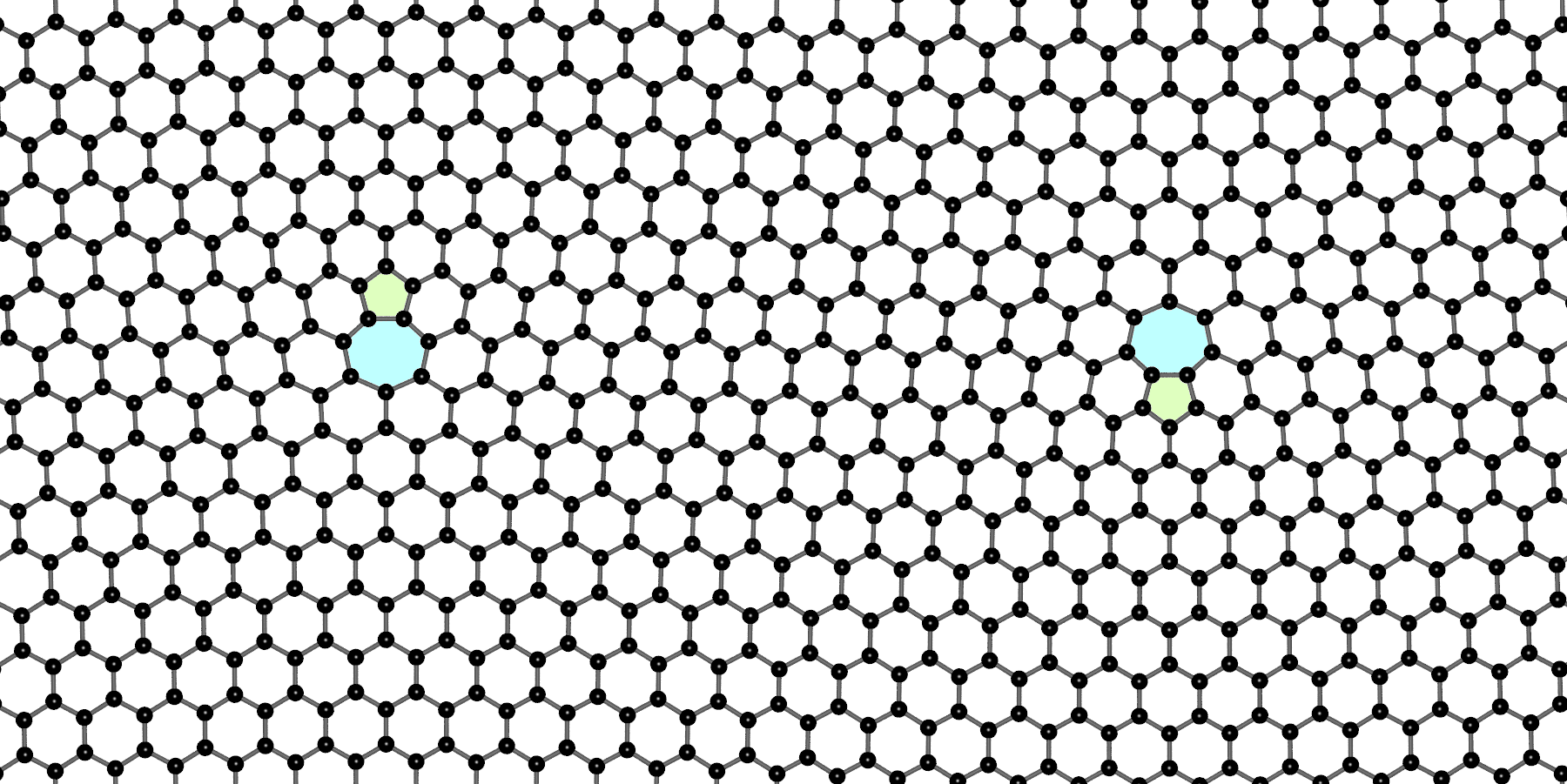}
\caption{(Color online) The system used for fitting PFC1, PFC3 and XPFC to DFT(2D). The depicted configuration consists of 780 atoms and has been relaxed using DFT(2D).}
\label{fig-DFT-fit-system}
\end{figure*}

The full periodic system used for fitting the energy scales of the PFC models to DFT is shown in Figure \ref{fig-DFT-fit-system}. The total width of the system is approximately 6.4 nm and it has 780 atoms. This is the only bicrystal system studied that was smaller than 10 nm in width. Table \ref{tab-atom-counts} gives the number of atoms in the grain boundary systems studied using different methods.

\begin{table}
\centering
\caption{The number of atoms in the grain boundary systems studied using PFC, DFT and MD. The upper limits for PFC models are approximate. The very large upper limits for XPFC and APFC are explained by large bicrystal widths employed at small tilt angles to eliminate elastic interactions and consequent slip of dislocations.}
\label{tab-atom-counts}
\begin{tabular}{l|c}
Model & Number of atoms \\ \hline
PFC1 & 412--\num{41600} \\
PFC3 & 412--\num{41600} \\
XPFC & 412--\num{166000} \\
APFC & \num{1140}--\num{5900000} \\
DFT & 412--\num{1040} \\
MD & 412--\num{12480} \\
\end{tabular} \\
\end{table}

\section{Calculation of elastic coefficients}
\label{sec-calculation-elastic}

Contribution of non-shearing elastic deformation to the free energy density of a two-dimensional system can be expressed as
\begin{equation}
\label{eq-elastic-free-energy}
\Delta f = \frac{C_{11}}{2}\left(\varepsilon_x^2+\varepsilon_y^2\right)+C_{12}\varepsilon_x\varepsilon_y,
\end{equation}
where $C_{11}$ and $C_{12}$ are stiffness coefficients, and $\varepsilon_x$ and $\varepsilon_y$ the two strain components \cite{ref-chaikin-lubensky}. We calculated the elastic free energy density landscape for single-crystalline PFC systems in the small deformation limit by applying varying combinations of uniform strain in the $x$ and $y$ directions. The stiffness coefficients were obtained from the least squares fit of Equation (\ref{eq-elastic-free-energy}) to the measured free energy density values. From $C_{11}$ and $C_{12}$, the elastic moduli: bulk modulus $B$, shear modulus $\mu$ and two-dimensional Young's modulus $Y_{2D}$, and Poisson's ratio $\nu$ can be solved \cite{ref-chaikin-lubensky} as
\begin{equation}
B = \frac{C_{11}+C_{12}}{2},
\end{equation}
\begin{equation}
\mu = \frac{C_{11}-C_{12}}{2},
\end{equation}
and
\begin{equation}
Y_{2D} = \frac{4B\mu}{B+\mu},
\end{equation}
where the bulk value of Young's modulus, $Y$, is given by dividing $Y_{2D}$ by the thickness of the monolayer, taken to be 3.35 Å. Last,
\begin{equation}
\nu = \frac{B-\mu}{B+\mu}.
\end{equation}

\section{Spectral defect detection}
\label{sec-sdd}

\begin{figure}
\includegraphics[width=0.48\textwidth]{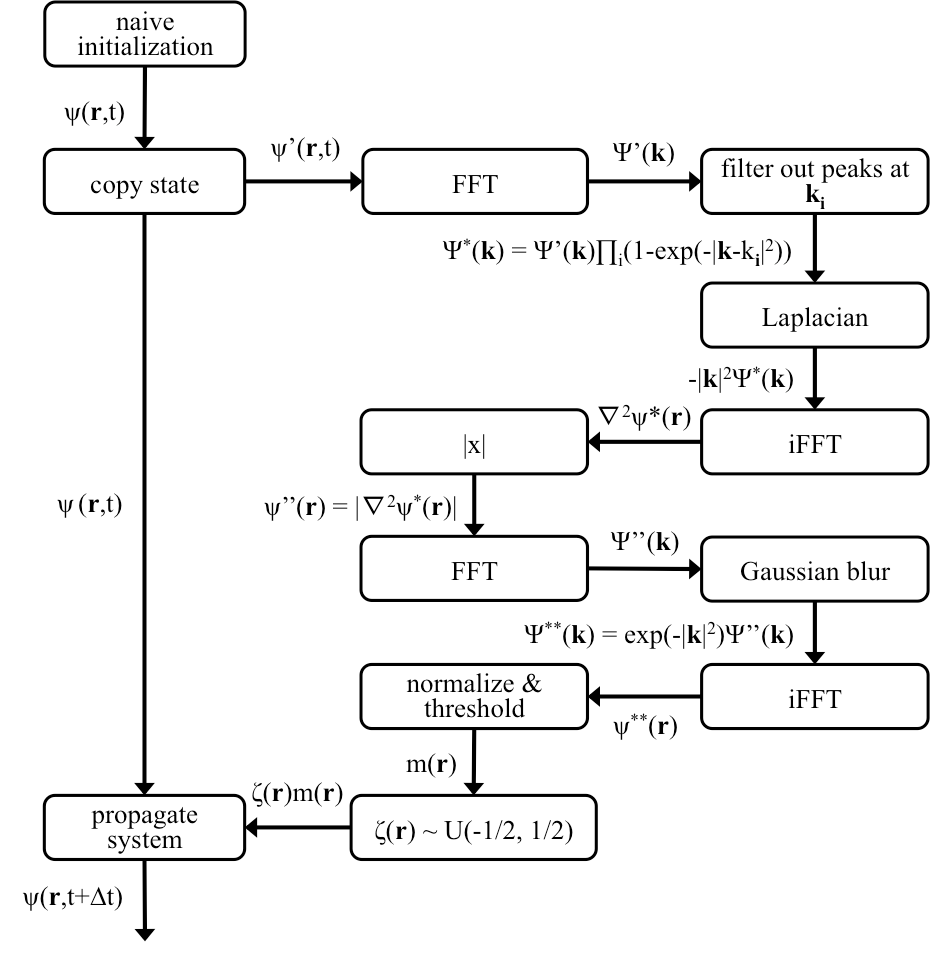}
\caption{Flowchart representation of the spectral algorithm for finding lattice defects. Position and time are denoted by $\boldsymbol{r}$ and $t$, respectively, while $\boldsymbol{k}$ is the Fourier space $\boldsymbol{k}$-vector. The binary mask is denoted by $m$ and $\zeta$ describes random noise sampled from a uniform distribution $U$.}
\label{fig-SDD}
\end{figure}

We developed a frequency filtering-based spectral defect detection (SDD) algorithm for focusing simulated annealing noise directly to lattice imperfections. Figure \ref{fig-SDD} presents a flowchart that illustrates the steps of the algorithm and gives the mathematical formulations thereof: Make a copy of the density field and compute its discrete Fourier transform. The bulk of the two bicrystal halves results in two sets of peaks in the amplitude spectrum, whose positions $\boldsymbol{k_i}$ are determined by the structure and rotation of the lattice. Filter out these peaks using smooth functions, e.g., Gaussians. Alternatively and especially for polycrystalline systems, instead of $\boldsymbol{k_i}$, filter out the full frequency bands $k = \left|\boldsymbol{k_i}\right|$. While still in Fourier space, take the Laplacian and then perform an inverse Fourier transform. Next, take the absolute value and apply some smoothing. We carried out this step by a Gaussian convolution in Fourier space. The steps described above result in a set of smooth bumps that are commensurate with the defects in the original density field. Then, normalize and threshold appropriately to obtain a binary mask $m=m\left(\boldsymbol{r}\right)$ that indicates the defected regions. Finally, use this mask to set the lattice imperfections to a disordered state or to focus annealing noise on them.

\bibliography{refs.bib}

\begin{thebibliography}{70}
\bibitem [{ref({\natexlab{b}})}]{ref-coloring}%
  \BibitemOpen
  \bibinfo {note} {The image is obtained via the following steps: normalize the
  density field, convolute with the kernel
  $\exp\left(-8\left(r-a\right)^2/a^2+\imath 6\theta\right)$ where $r$ and
  $\theta$ are the polar coordinates and $a$ is the bond length, multiply with
  the normalized squared, smooth out the atomic structure, and map the phase
  and amplitude linearly to hue and brightness, respectively. An alternative
  technique is introduced in detail in Reference
  \cite{ref-wavelet}.}\BibitemShut {Stop}%
\end{thebibliography}

\end{document}